\documentclass{scrartcl}

\usepackage[utf8]{inputenc}
\usepackage[T1]{fontenc}
\usepackage{authblk}

\usepackage{graphicx}
\usepackage{amsmath,amssymb,amsthm}
\usepackage{esint} 
\usepackage{enumitem,color}
\usepackage{comment}
\usepackage{hyperref}

\usepackage{xstring}

\theoremstyle{definition}

\theoremstyle{remark}

\newcommand{\fig}[2]{\includegraphics[width=#1\textwidth]{#2}}

\newcommand{\N}{\mathbb{N}}
\newcommand{\Z}{\mathbb{Z}}
\newcommand{\R}{\mathbb{R}}

\newcommand{\A}{\boldsymbol{A}}
\newcommand{\B}{\boldsymbol{B}}
\newcommand{\C}{\boldsymbol{C}}
\newcommand{\X}{\boldsymbol{X}}
\newcommand{\M}{\boldsymbol{M}}
\newcommand{\btwo}{\boldsymbol{b_2}}

\title{Exponents for Hamiltonian paths on random bicubic maps and KPZ}
\author[1,2]{Philippe Di Francesco\thanks{philippe.di-francesco@ipht.fr}}
\author[1]{Bertrand Duplantier\thanks{bertrand.duplantier@ipht.fr}}
\author[1] {Olivier Golinelli\thanks{olivier.golinelli@ipht.fr}}
\author[1]{Emmanuel Guitter\thanks{emmanuel.guitter@ipht.fr}}
\affil[1]{{\normalsize Université Paris-Saclay, CEA, CNRS, Institut de physique
    théorique, 91191, Gif-sur-Yvette, France}}
\affil[2]{{\normalsize Department of Mathematics, University of Illinois, Urbana, IL 61821,
USA}}

\date{\today}

\begin{document}
\maketitle 
\begin{abstract}
We evaluate the configuration exponents of various ensembles of Hamiltonian paths 
drawn on random planar bicubic maps. These exponents are estimated from the extrapolations
of exact enumeration results for finite sizes and compared with their theoretical predictions based
on the KPZ relations, as applied to their regular counterpart on the honeycomb lattice. 
We show that a naive use of these relations does not reproduce
the measured exponents but that a simple modification in their application may possibly correct
the observed discrepancy. We show that a similar modification is required to reproduce 
via the KPZ formulas some exactly known exponents for the problem of unweighted fully packed loops
on random planar bicubic maps. 
  \end{abstract}

\section{Introduction}
\label{sec:intro}
The aim of this paper is to evaluate a number of exponents characterizing the asymptotic enumeration of 
various configurations of Hamiltonian paths on random planar bicubic maps. Recall that a 
\emph{planar map} is a connected graph drawn on the two-dimensional (2D) sphere (or equivalently on 
the plane) without edge crossings, and considered up to continuous deformations. A map is characterized by its 
vertices and edges, inherited from the underlying graph structure, and by its  \emph{faces} 
which result from the embedding and all have the topology of the disk. The map is called \emph{bicubic} if (i) it is cubic, i.e., all 
its vertices have degree $3$ (i.e., have $3$ incident half-edges) and (ii) these vertices are colored in black and white 
so that any two adjacent vertices have different colors. A \emph{Hamiltonian} path is a self-avoiding path along 
the edges of the map which visits all the vertices of the map. If the path is closed, it is called a 
\emph{Hamiltonian cycle}. 

\begin{figure}
  \centering
  \fig{.7}{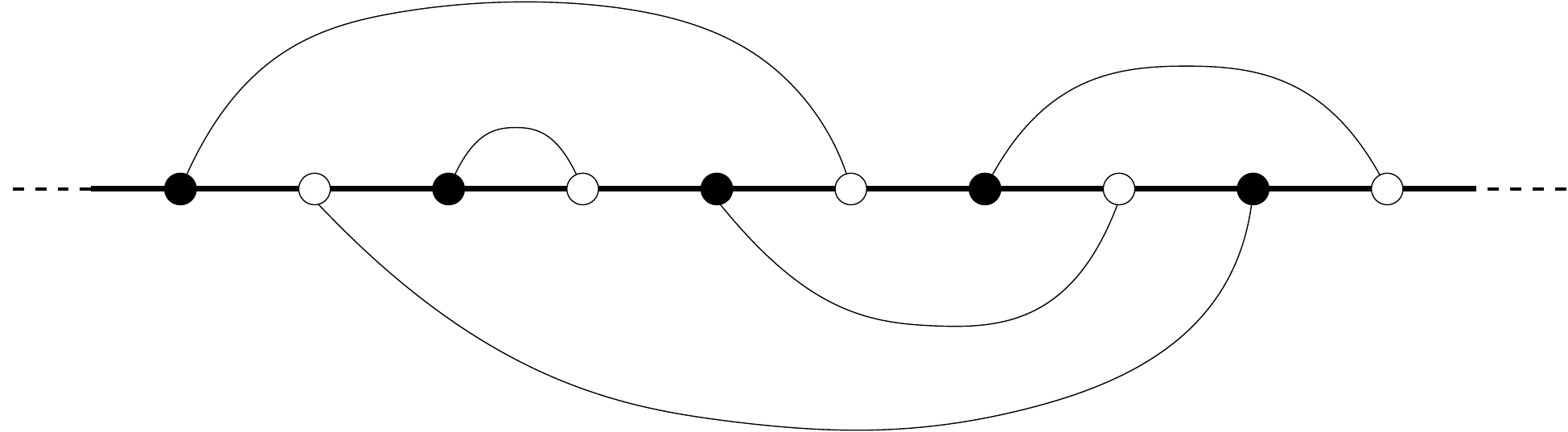}
   \caption{\small Representation of a (rooted) Hamiltonian cycle on a planar bicubic map as a system of non-crossing arches 
   linking black and white vertices whose color alternate along a straight line.}
  \label{fig:arches}
\end{figure}

The problem of Hamiltonian cycles on random bicubic maps was first considered in  \cite{GKN99} 
where it was conjectured that its scaling limit corresponds to 2D quantum gravity coupled to a 
conformal field theory (CFT) with central charge $c=-1$. In particular, the number  $z_N$ of configurations of 
planar bicubic maps with  $2N$ vertices endowed with a Hamiltonian cycle and a marked visited edge, 
called the \emph{root} edge, was predicted in \cite{GKN99} to behave, at large $N$, as
 \begin{equation}
 z_N\sim \hbox{const.}\ \frac{\mu^{2N}}{N^{2-\gamma}}
 \label{eq:zN}
 \end{equation}
 with an exponential growth rate estimated numerically as $\hbox{Log}(\mu^2)~\sim2.313$ and with the somewhat 
 nontrivial exponent
 \begin{equation}
 \gamma=-\frac{1+\sqrt{13}}{6}\ .
 \label{eq:gammaval}
 \end{equation}
By cutting the Hamiltonian cycle at the level of its root edge and stretching it into a straight line, 
a configuration may be drawn in the plane as a simple infinite line  with $2N$ alternating
black and white vertices, completed by non-crossing arches linking each black vertex to a white one,
either above or below the infinite line, see Figure~\ref{fig:arches} for an example. Despite this simple 
representation, no exact expression for $z_N$ or its asymptotic equivalent is known so far and the results 
of \cite{GKN99} remain mathematically a conjecture.

\bigskip
In the present paper, we address the question of evaluating a number of other exponents, similar to $\gamma$
and characterizing more involved Hamiltonian path configurations with possible valency and/or occupation defects. 
In our study, we will be led to consider a generalization of the Hamiltonian cycle problem to the so-called 
FPL$(n)$ model on bicubic maps, where FPL stands for \emph{Fully Packed Loops}. Configurations of the FPL$(n)$ model
now consist of an arbitrary number of \emph{loops} which are closed paths drawn along the edges of the underlying bicubic
map, so that loops are both self- and mutually avoiding and each vertex of the map is visited by a loop.
Each loop receives the weight $n$, with $n$ a real number between $0$ and $2$. Three values of
$n$ are of particular interest: the case $n=2$ describes unweighted \emph{oriented} loops, the
case $n=1$ that of unweighted loops, while taking the limit $n\to 0$ allows us to recover the original 
Hamiltonian cycle problem with a single loop.

A strategy to explore the asymptotics of Hamiltonian cycles, or more generally that of the  FPL$(n)$ model
on random bicubic maps consists in using the general connection which links 2D conformal field
theories in random geometry to their classical (Euclidean) counterpart via the so-called \emph{KPZ formulas} \cite{KPZ88}.
These formulas act as a translation tool giving global configuration exponents for the random map problem from the value of the associated
critical correlation exponents (also called classical dimensions) in the corresponding regular lattice problem.
Note that this ``KPZ strategy'' was carried out successfully in \cite{DGG00} and \cite{DGJ00} (see also \cite{DFG05}) to identify various configuration exponents 
for \emph{meanders}, a related combinatorial problem with now \emph{two} intertwined fully packed loops drawn on random tetravalent planar maps.

\bigskip
The regular infinite bicubic map is nothing but the honeycomb lattice. The FPL$(n)$ model on the honeycomb lattice
was considered in \cite{BN94} and many exact results were obtained by various approaches such
as Bethe Ansatz techniques \cite{BSY94} or more heuristic Coulomb Gas (CG) methods \cite{KdGN96},
see also \cite{DEI19}. We can therefore deduce from these results a number of KPZ-induced configuration exponents for the random
map problem. In the case $n=0$, we can then compare the predicted values for these exponents with their numerical estimates obtained 
from the extrapolation of \emph{exact enumeration results for finite $N$}, as was done with success in \cite{GKN99} 
for the exponent $\gamma$ above. As it will appear, the ``naive'' KPZ approach, based on a direct application of the KPZ formulas,
 does not lead to satisfactory results for $n=0$. Still it seams that the observed mismatch with numerical estimates might be corrected 
 if, before applying the KPZ formulas, we slightly modify the expression of the critical correlation exponents by a  
 simple extra ``normalization'' procedure involving a single parameter $\alpha$. As we shall see, a similar $\alpha$-corrected KPZ 
procedure is required in the case $n=1$ when comparing KPZ-induced configuration exponents to exactly known results for specific observables.

\bigskip
The paper is organized as follows:
 Section~\ref{sec:CG} presents a number of known results for the FPL$(n)$ model on the honeycomb lattice: after recalling
 its Coulomb Gas description in Section~\ref{sec:CGgeneral}, we give the expression for various critical exponents 
 corresponding to vortex-antivortex correlations in Section~\ref{sec:vortex}. Specific examples of these correlations 
 and their meaning in terms of loops are discussed in Section~\ref{sec:examples}. We then turn in Section~\ref{sec:FPL} to the 
 coupling of the FPL$(n)$ model to gravity. After discussing in Section~\ref{sec:bicubic} the specificity of bicubic maps
 in connection with foldable triangulations, we recall in Section~\ref{sec:KPZ} the general KPZ relations for the coupling to gravity
 of a 2D conformal theory. We also discuss in Section~\ref{sec:scalinglimit} the expected limits of
 both the standard O$(n)$ model and the FPL$(n)$ model in terms of Schramm-Loewner Evolution (SLE) and Liouville
 quantum gravity (LQG). We then use in Section~\ref{sec:none} an 
 equivalence between the FPL$(1)$ model on bicubic maps and
 a particular instance of the 6-vertex model on tetravalent maps to obtain, in the case $n=1$, the configuration exponents for a family
 of vortex-antivortex correlations. We note that, quite surprisingly, the direct application of the KPZ formulas does not reproduce these results
 but that the observed discrepancy is easily cured in this case if we allow for a slight modification of the classical dimensions
 before applying the KPZ formulas.  Our main results are presented in Sections~\ref{sec:numerics} and \ref{sec:comparison}:
 Section~\ref{sec:numerics} deals with the exact numerical enumeration of configurations of Hamiltonian paths on bicubic maps
 with finite sizes and possible defects. After discussing in Section~\ref{sec:enumeration} our enumeration methods, we present in
 Section~\ref{sec:enumerate} our enumeration results for various ensembles of configurations with maximal sizes ranging from 
 $N=16$ up to $N=28$. These results are then used in Sections~\ref{sec:growthrate}
and \ref{sec:exponents} respectively to estimate the asymptotic exponential growth rate and the configuration exponent for each of
the configuration ensembles at hand. The comparison with the KPZ predictions is then discussed in Section~\ref{sec:comparison}.
Again we observe a mismatch with the numerical estimates and we present a tentative $\alpha$-corrected KPZ procedure
which seems to resolve this discrepancy. We gather a few remarks in Section~\ref{sec:discussion}. In particular, we explore the possible meaning
of the parameter $\alpha$. Appendix~\ref{appendix:cubic}
discusses configuration exponents for Hamiltonian paths drawn on (non-necessarily bicolorable) planar cubic maps 
and shows the validity of the naive KPZ approach in this case. Appendix~\ref{appendix:results} 
presents the complete list of our exact enumeration results for the various Hamiltonian path configurations 
on bicubic maps that we have studied.

\section{Coulomb Gas description of the FPL$\boldsymbol{(n)}$ model on the honeycomb lattice}
\label{sec:CG}

\subsection{General theory}
\label{sec:CGgeneral}
\begin{figure}
  \centering
  \fig{.42}{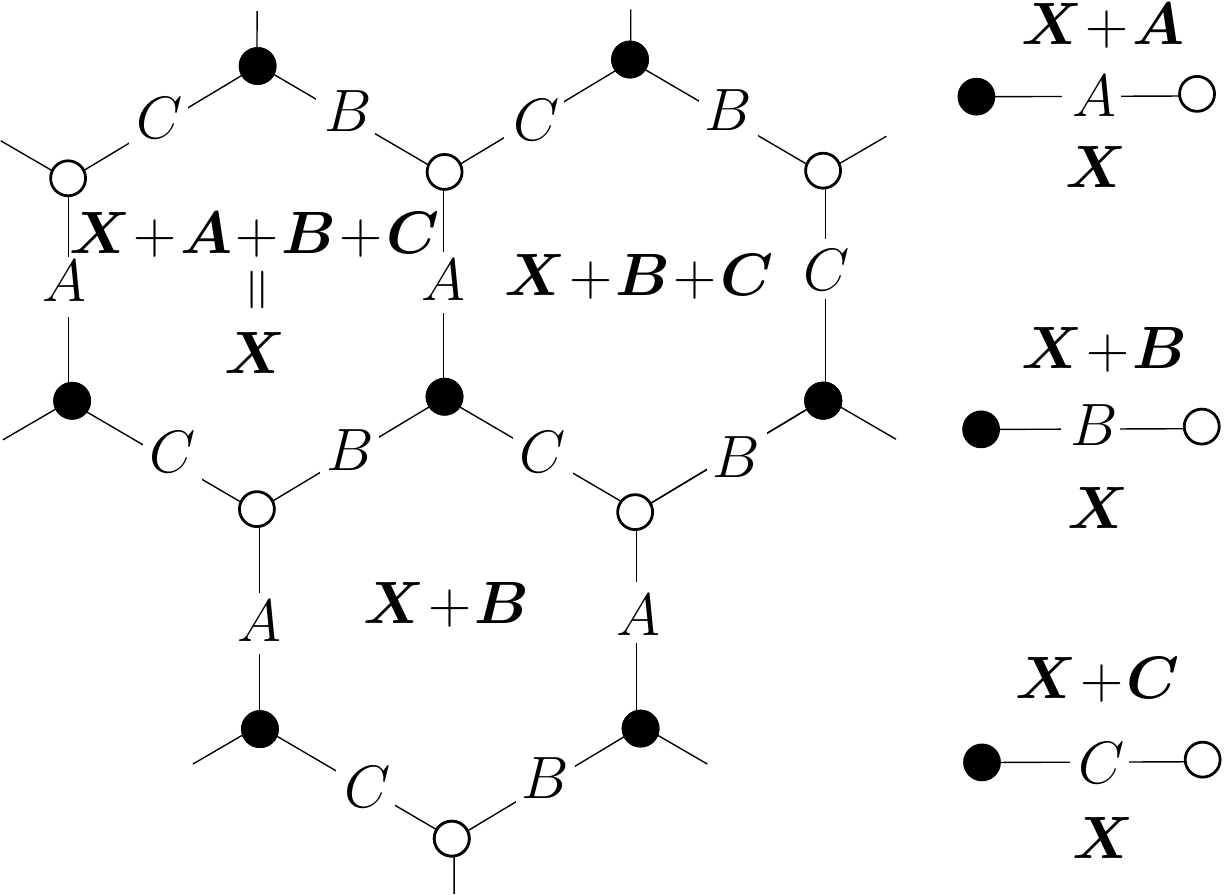}\hskip 1.5cm \fig{.42}{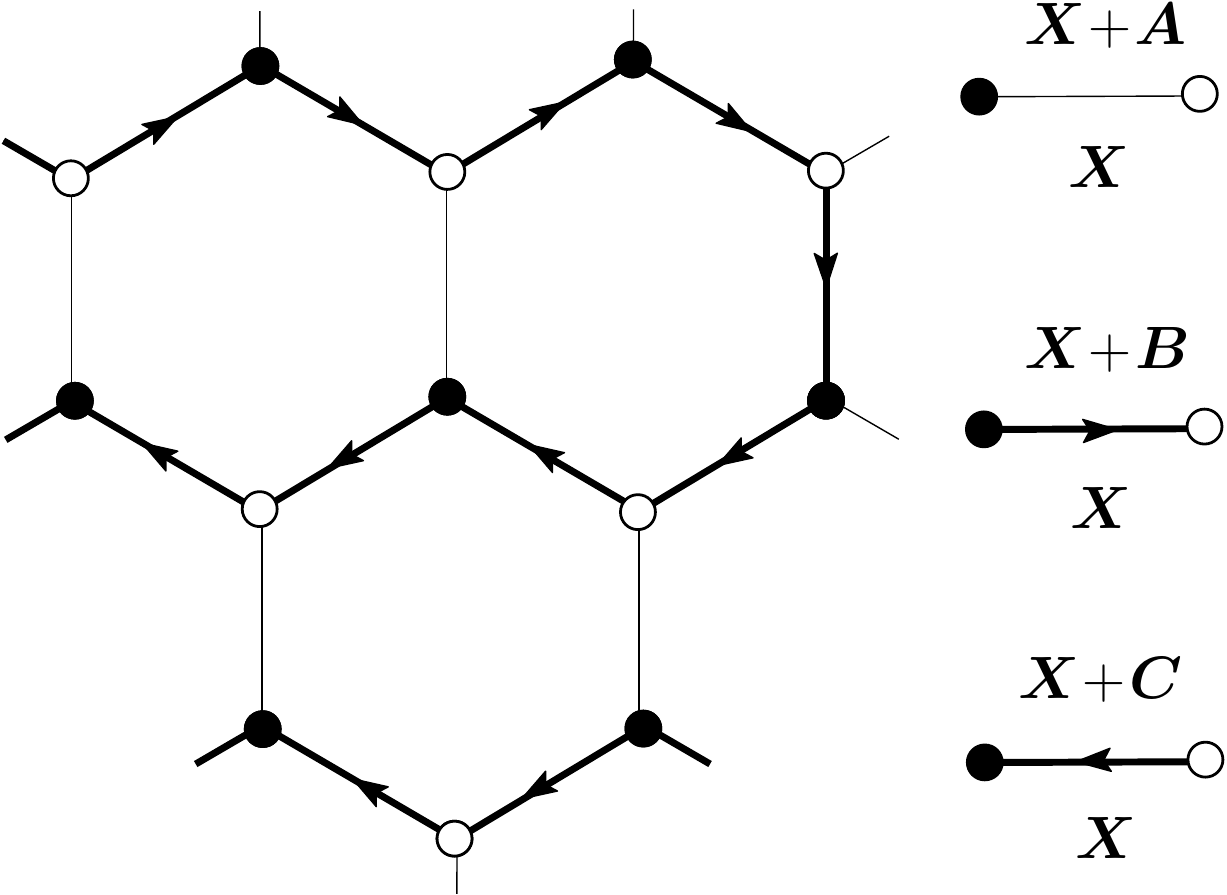}
  \caption{\small Left: an edge $3$-coloring configuration of a portion of honeycomb lattice with colors $A$, $B$ and $C$.
  Right: the associated configuration of fully packed oriented loops. The $B$- and $C$-colored edges form the oriented loops, 
  while the $A$-edges correspond to the unvisited edges. For each representation, we indicated the variation
  of the 2D height variable $\X$ when crossing an edge. }
  \label{fig:honeycomb}
\end{figure}

Following \cite{KdGN96}, we start with the description of the FPL$(2)$ model on the honeycomb lattice which,
as discussed above, corresponds to configurations of fully packed oriented loops.
As displayed in Figure~\ref{fig:honeycomb}, a configuration can be alternatively described as a $3$-coloring of
the edges of the lattice by colors $A$, $B$ and $C$, so that the three edges incident to any vertex be of
different colors. It is indeed easily seen that, for such a $3$-coloring, the $B$- and $C$-colored edges form closed 
loops of alternating $B$- and $C$-edges visiting all the vertices of the lattice, while the $A$-edges correspond to the unvisited edges.
Orienting the visited edges from their black to their white incident vertex 
for $B$-edges, and  from their white to their black incident vertex for $C$-edges
induces a well-defined orientation for each loop. Changing the orientation of a loop 
simply corresponds to interchanging the $B$- and $C$-edges along it.

We may finally transform the FPL$(2)$ configurations into a ``height-model'' by assigning
to each hexagonal face a \emph{two-dimensional} height $\X\in\R^2$ whose variation $\Delta \X$ between neighboring
faces depends on the nature of their separating edge, with the dictionary of Figure~\ref{fig:honeycomb}.
In the $3$-coloring language, we have $\Delta \X=\A$ (resp. $\B$, $\C$) if the crossed edge is
of color $A$ (resp. $B$, $C$) and traversed with the incident black vertex on the left.
Making a complete turn around any vertex of the honeycomb lattice implies the constraint
$\A+\B+\C=\boldsymbol{0}$ so that $\X$ is indeed two-dimensional.

\begin{figure}
  \centering
  \fig{.2}{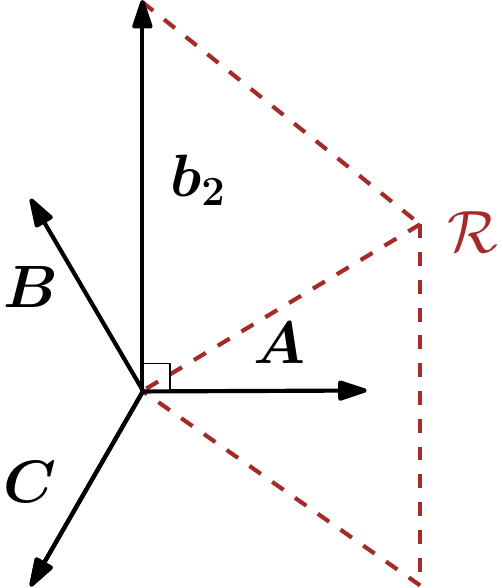}
  \caption{\small The two-dimensional vectors $\A$, $\B$, $\C$ and $\btwo$ and a portion of the $\mathcal R$ lattice.}
  \label{fig:frame}
\end{figure}

Following \cite{KdGN96}, we make the following symmetric choice of vectors, see Figure~\ref{fig:frame}:
\begin{equation}
\A:=\left(\frac{1}{\sqrt{3}}, 0 \right)\ , \quad
\B:=\left(-\frac{1}{2\sqrt{3}},  \frac{1}{2}\right)\ , \quad
\C:=\left(-\frac{1}{2\sqrt{3}}, -\frac{1}{2} \right)\ ,
\end{equation}
so that $\vert\A\vert=\vert\B\vert=\vert\C\vert={1}/{\sqrt{3}}$
and $\A\cdot\B=\B\cdot\C=\C\cdot\A=-{1}/{6}$.
The height variable $\X$ takes its values within the triangular lattice $\mathcal{T}:=\Z\B+\Z\C$, with mesh size ${1}/{\sqrt{3}}$. 
We also define the ``repeat lattice'' as the sub-lattice of $\mathcal{T}$ given by  
\begin{equation}
\mathcal{R}:=\Z\, (\A-\B)+\Z\, (\A-\C)
\end{equation}
which is a triangular lattice of mesh size $1$ (see Figure~\ref{fig:frame}). It is such that two pieces of lattice whose values of the 2D height differ globally by an element 
of $\mathcal{R}$ describe the same coloring arrangement, hence the same loop configuration environment.
 
In  \cite{KdGN96}, it is claimed that the model may then be described at large scale by a coarse-grained variable
$\boldsymbol{\Psi}(x)=\langle \X \rangle$  suitably averaged in the vicinity of the point $x$ of the underlying lattice, 
where $\boldsymbol{\Psi}$ is governed by the free field action 
\begin{equation}
\mathcal{A}_{CG}=\pi\, g\, \int d^2 x \left(\nabla\boldsymbol{\Psi}\right)^2
\end{equation}
with $g=1$ (with our choice of normalization for $\A,\B,\C$). The variable $\boldsymbol{\Psi}$ is defined modulo $\mathcal{R}$, i.e.,
is an element of $\R^2/\mathcal{R}$ via the equivalence relation in $\R^2$:
\begin{equation}
\label{eq:Rshift}
\boldsymbol{\Psi}\equiv \boldsymbol{\Psi'} \iff \boldsymbol{\Psi}- \boldsymbol{\Psi'}\in \mathcal{R}\ .
\end{equation}
Since the loops follow the $B$- and $C$-colored edges, the color $A$ on the one hand and the colors 
$B$ and $C$ on the other hand play very different roles in the description of loop observables. It 
is then convenient to introduce the vector 
\begin{equation}
\btwo:= \B-\C=\left(0,1\right)\ ,
\end{equation}
so that $\vert\btwo\vert=1$ and $\btwo\cdot \A=0$, and to work in the \emph{orthogonal basis} $(\A,\btwo)$, see Figure~\ref{fig:frame}.
We thus write
\begin{equation}
\boldsymbol{\Psi}=\psi_1 \A+\psi_2\btwo\ .
\end{equation}
For a fixed $n$ ($0\leq n\leq 2$), the wanted weight $n$ per loop in the  FPL$(n)$ model is obtained by introducing
\emph{local weights} accounting for the left or right nature of the turns of the (oriented) loops at each vertex. 
At large scales, this new weight results into a modified Coulomb Gas action (see \cite{KdGN96,DEI19} for details)
 \begin{equation}
\mathcal{A}_{CG}=
 \int d^2 x\left\{ \pi\, g\, \left(\frac{1}{3}\left(\nabla{\psi_1}\right)^2+\left(\nabla{\psi_2}\right)^2\right)
+\frac{1}{2}\hbox{i}\, e_0\, \psi_2 \, R+: e^{4\mathrm{i}\pi \psi_2}:\right\}
\label{eq:action}
\end{equation}
where $R$ is the (local) scalar curvature of the underlying lattice\footnote{For instance, if the model is defined on a cylinder by taking periodic conditions
in one direction, the scalar curvature is concentrated at both ends of the cylinder.}, and with now
\begin{equation}
\label{eq:gdef}
g=1-e_0=1-\frac{1}{\pi}\arccos\left(\frac{n}{2}\right)\ .
\end{equation}
The last term in the action is the most relevant perturbation which allows one to fix $g$ by demanding that it be marginal \cite{DEI19}.
At this stage, it is important to note that the action is such that the two components
$\psi_1$ and $\psi_2$ of the field $\boldsymbol{\Psi}$ are decoupled. The
component $\psi_1$ along $\A$ is governed by a simple free field action, while
the component $\psi_2$ along $\btwo$ is governed by a usual \emph{one-dimensional} CG action, similar
to that obtained from the SOS reformulation of a dense O$(n)$ model (i.e., without the constraint that
each vertex is visited by a loop).   
Still, a coupling between the two directions $\A$ and $\btwo$ arises when
we deal with the operator spectrum of the FPL$(n)$ model, as the defect configurations
must be consistent with the condition \eqref{eq:Rshift}.
Finally, the central charge of the FPL$(n)$ model on the honeycomb lattice is given by
\begin{equation}
c_{\mathrm{fpl}}(n)=1+c_{\mathrm{dense}}(n)=2-6 \frac{(1-g)^2}{g}
\label{eq:cc}
\end{equation}
where, in the middle expression, the first term $1$ is the central charge for the free scalar field $\psi_1$ 
and the second term is the usual central charge
\begin{equation}
c_{\mathrm{dense}}(n)=1-6 \frac{(1-g)^2}{g}
\label{eq:cdense}
\end{equation} of a dense O$(n)$ model
(as obtained for instance from its one-dimensional CG description \cite{zbMATH03959008}). For $n= 0$
($g= \frac{1}{2}$), this yields a total central charge $c_{\mathrm{fpl}}(0)=1-2=-1$.

\subsection{Exponents for vortex-antivortex correlations}
\label{sec:vortex}
The operator spectrum of the FPL$(n)$ model on the honeycomb lattice is discussed in details in \cite{KdGN96}. Of particular
interest are the so-called \emph{vortex operators} characterized, in the CG language, by 
their 2D \emph{magnetic charge} $\M\in \mathcal{R}$. An operator with magnetic charge $\M$ corresponds to the insertion
of a dislocation at a given lattice vertex, i.e., a topological defect $\delta \X= \M$ in the 2D height 
when going counterclockwise around that vertex. Several defects must be introduced simultaneously so that the total magnetic charge 
is zero (this guarantees that the 2D height remains well defined at infinity) to keep a finite free energy cost. For
instance,  the vortex-antivortex correlation corresponds to inserting a vortex of charge $\M$ and one of charge 
$-\M$ at two fixed vertices distant by $r$ in the honeycomb lattice.  
For
\begin{equation}
\M=j(\A-\B) +k(\A-\C)\in \mathcal{R} , \quad {j,k\in\Z}\ ,
\end{equation}
the change $\Delta F$ of free energy induced by the introduction of the $\M/-\M$ defect is expected to behave at large $r$ as 
$e^{-\Delta F}\sim r^{-4 h_{\M}}$ with exponent \cite{KdGN96}
\begin{equation}
h_{\M}=\frac{g}{4}\left(j^2+k^2+j\, k\right) -\frac{(1-g)^2}{4g}\left(1-\delta_{j,k}\right)\ .
\label{eq:valexpo}
\end{equation}
Using instead coordinates in the orthogonal basis $(\A,\btwo)$, i.e., writing $\M=\frac{3}{2}(j+k)\A+\frac{1}{2}(k-j)\btwo$,
we may recast the above result into
\begin{equation}
\label{eq:valexp}
h_{\M}=\frac{g}{12}\phi_1^2 +\frac{g}{4}\left(1-\delta_{\phi_2,0}\right)\left(\phi_2^2-\left(1-g^{-1}\right)^2\right)
\qquad \hbox{for}\quad \M=\phi_1 \A+\phi_2\btwo \ ,
\end{equation}
where the coordinates $\phi_1$ and $\phi_2$ are now integers or half-integers satisfying  $\phi_1\in \frac{3}{2}\Z$, $\phi_2\in \frac{1}{2}\Z$ and
$\phi_1+\phi_2\in \Z$. Note that, as a consequence of the decoupled form \eqref{eq:action} of the action, the expression above for $h_{\M}$ 
is naturally split into two terms: a first contribution depending on the coordinate $\phi_1$ along $\A$ only
and a second contribution involving the coordinate $\phi_2$ along $\btwo$ only.
  
\subsection{Examples}
\label{sec:examples}
Let us illustrate the result \eqref{eq:valexp} in the case $n=0$ and for a few values of the magnetic charge $\M$. 
For $n=0$ ($g=\frac{1}{2}$), Equation~\eqref{eq:valexp} reduces to 
\begin{equation}
\label{eq:valexp0}
h_{\M}(n=0)=\frac{1}{24}\phi_1^2 +\frac{1}{8}\left(1-\delta_{\phi_2,0}\right)\left(\phi_2^2-1\right)\ .
\end{equation}
\begin{figure}
  \centering
  \fig{.7}{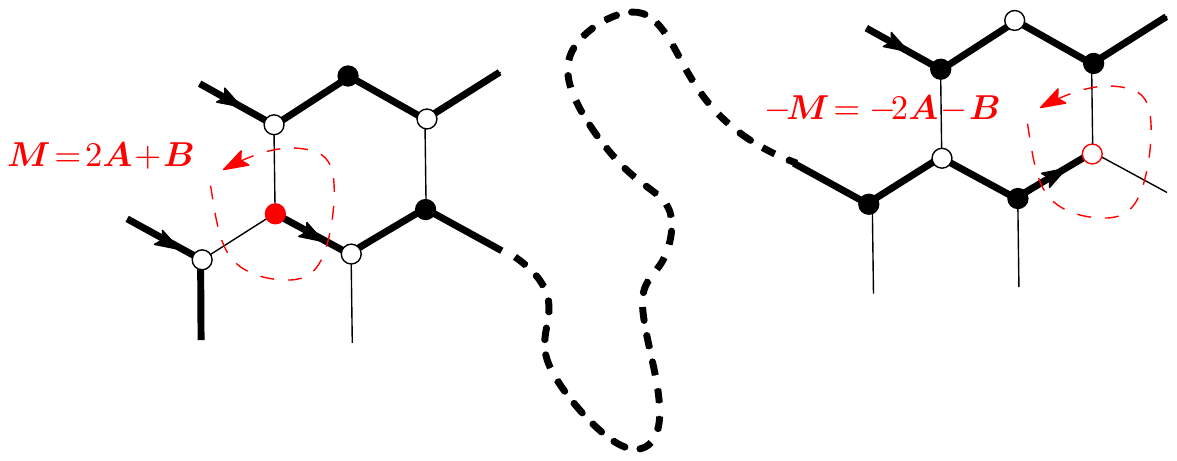}
  \caption{\small Schematic representation of a pair of defects with magnetic charges $\pm\M$ for $\M=2\A+\B$ in the case of arbitrary $n$. This corresponds to fully packed
  loop configurations containing an open oriented path starting at a black vertex and ending at a white one. If these vertices are at distance
  $r$ apart from each other on the honeycomb lattice, the change $\Delta F$ of free energy due to the defects is expected to behave 
  as $e^{-\Delta F}\sim r^{-4 h_{\M}}$.}
  \label{fig:BtwoA}
\end{figure}
\paragraph{The case $\M=\B+2\A=\frac{3}{2}\A+\frac{1}{2}\btwo$.}
A vortex with magnetic charge $\B+2\A$ corresponds to a black vertex surrounded by two $\A$'s and one $\B$ which, in the loop language, corresponds to a black vertex from which a path originates (see Figure~\ref{fig:BtwoA}). The corresponding antivortex, of charge  
$-\B-2\A$ corresponds to the end of this path at a further apart white vertex. For $n\to 0$, this vortex-antivortex correlation
therefore enumerates configurations with a single \emph{open} path which is fully packed, i.e., visits all the vertices
of the lattice, and with prescribed (black) starting and (white) ending points at distance $r$ from each other.
From \eqref{eq:valexp0}, we have
\begin{equation}
h_{\B+2\A}(n=0)=\frac{1}{24}\left(\frac{3}{2}\right)^2+\frac{1}{8}\left(\left(\frac{1}{2}\right)^2-1\right)=0\ ,
\end{equation}
meaning that the free energy cost induced by having remote endpoints for the path tends to a constant for large $r$.
A possible explanation for this result is that, for $n\to 0$, the length of the Hamiltonian path joining the defects
is independent of $r$ and depends only on the size of the underlying lattice. 
\begin{figure}
  \centering
  \fig{.7}{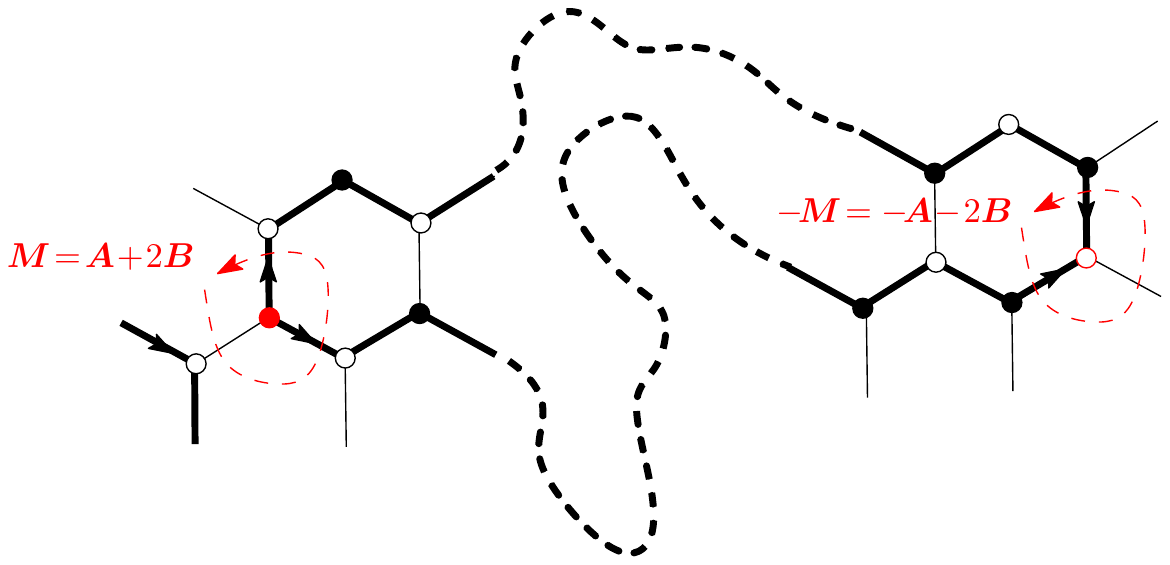}
  \caption{\small Schematic representation of a pair of defects with magnetic charges $\pm\M$ for $\M=\A+2\B$
  in the case of arbitrary $n$. This corresponds to fully packed
  loop configurations with two open oriented paths starting at a black vertex and meeting again at a (black or white) vertex at distance 
  $r$ apart.}
  \label{fig:AtwoB}
\end{figure}

\paragraph{The case $\M=\A+2\B=\btwo$.}
A vortex with magnetic charge $\A+2\B$ corresponds to two paths originating from the same black vertex (see Figure~\ref{fig:AtwoB}),
and ending at the white vertex\footnote{We may also put the antivortex at a black vertex, now with magnetic charge
$\A+2\C$ so that the total charge is $\A+2\B+\A+2\C=\boldsymbol{0}$.} where we put the antivortex  $-\A-2\B$. By changing the orientation of one of the paths,
the concatenation of the two paths forms a well-oriented loop: the vortex-antivortex correlation
therefore enumerates fully packed loop configurations for which two prescribed vertices
on the honeycomb lattice at distance $r$ from each other belong to the $\emph{same}$ loop. For $n\to0$, this is always the case
since the configuration is made of a single cycle. This is consistent with the value 
\begin{equation}
h_{\A+2\B}(n=0)=\frac{1}{8}\left((1)^2-1\right)=0\ .
\end{equation}

\begin{figure}
  \centering
  \fig{.7}{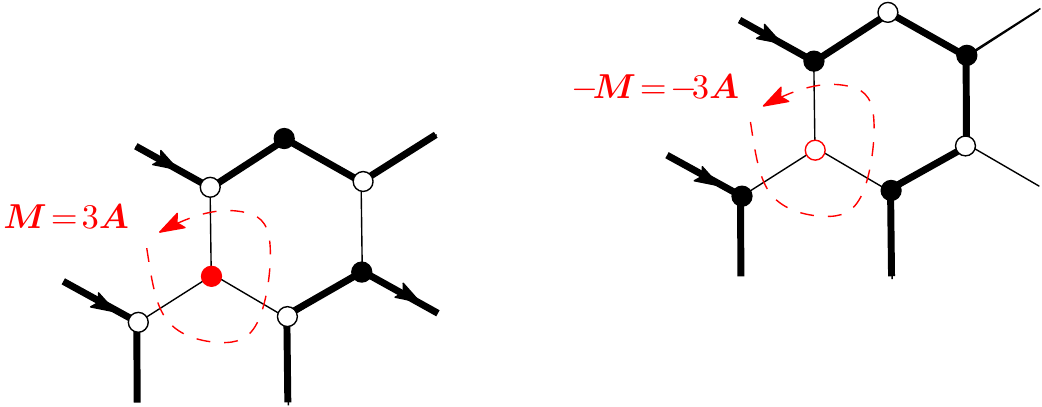}
  \caption{\small Schematic representation of a pair of defects with magnetic charges $\pm\M$ for $\M=3\A$. This corresponds to fully packed
  loop configurations with two unvisited vertices, one of each color, at given distant  
  $r$ from each other.}
  \label{fig:threeA}
\end{figure}
\paragraph{The case $\M=3\A$.}
This corresponds to having a black and a white vertex at distance $r$ apart which are not visited by a loop.
For $n\to 0$, we then have a unique fully packed loop visiting all vertices but two (see Figure~\ref{fig:threeA}).
We find
 \begin{equation}
h_{3\A}(n=0)=\frac{1}{24}(3)^2=\frac{3}{8}\ ,
\end{equation}
meaning that the number of configurations decays  as $r^{-3/2}$ at large $r$. 
Since $h_{3\A}<1$, 
such a defect corresponds to a relevant perturbation. This agrees with the fact that the 
fully packed loop fixed point (here for $n=0$) is unstable with respect to the creation of ``empty'' vertices:
giving these vertices a finite chemical potential drifts the model toward 
a new fixed point, that of the dense $\mathrm{O}(n)$ model, with central charge $c_{\mathrm{dense}}(n)$ as in \eqref{eq:cdense} \cite{zbMATH03959008}.

\section{Coupling to gravity}
\label{sec:FPL}

\subsection{A word on bicubic maps}
\label{sec:bicubic}

The vertices of the honeycomb lattice all have degree $3$. As such, the honeycomb lattice may be viewed
as the regular lattice associated with random \emph{cubic} maps. However, as opposed to the honeycomb lattice,
an arbitrary cubic map is not vertex bicolorable in general. The existence of a bicoloring of the vertices
in black and white was crucial when defining the 2D height $\X$ coding for a fully packed (oriented) loop
configuration. Without this coloring, it is not possible to distinguish between the two sides of an unvisited edge: this
forces us to set $\A=\boldsymbol{0}$ and $\B=-\C$ accordingly, leading to a height $\X$ which is one-dimensional
only.
A similar reduction of the dimension from $2$ to $1$ occurs on the honeycomb lattice itself if we allow for the presence
of unvisited vertices with a finite chemical potential, as it imposes $3\A=\boldsymbol{0}$. 
The corresponding ``densely packed'' O$(n)$ model is much simpler than the FPL$(n)$ model and  
corresponds to a conformal theory of central charge $c_{\mathrm{dense}}(n)$ as in \eqref{eq:cdense}, with $g$ as in 
\eqref{eq:gdef}, which can be described at large scales by standard one-dimensional CG techniques 
\cite{zbMATH03959008}. The FPL$(n)$ model, when defined  on arbitrary random cubic maps, is therefore expected to be described by 
a conformal theory of reduced central charge $c=c_{\mathrm{dense}}(n)$ coupled to gravity. For $n=0$, this yields $c=-2$,
a result which can directly be verified by an exact enumeration of the configurations 
(see Appendix \ref{appendix:cubic}).

\medskip 
To get a non-trivial FPL$(n)$ model with the augmented central charge $c_{\mathrm{fpl}}(n)$ of \eqref{eq:cc}, we therefore need to impose that the vertices of the random cubic map be 
\emph{bicolored}, namely that the random map be \emph{bicubic}.
\begin{figure}
  \centering
  \fig{.75}{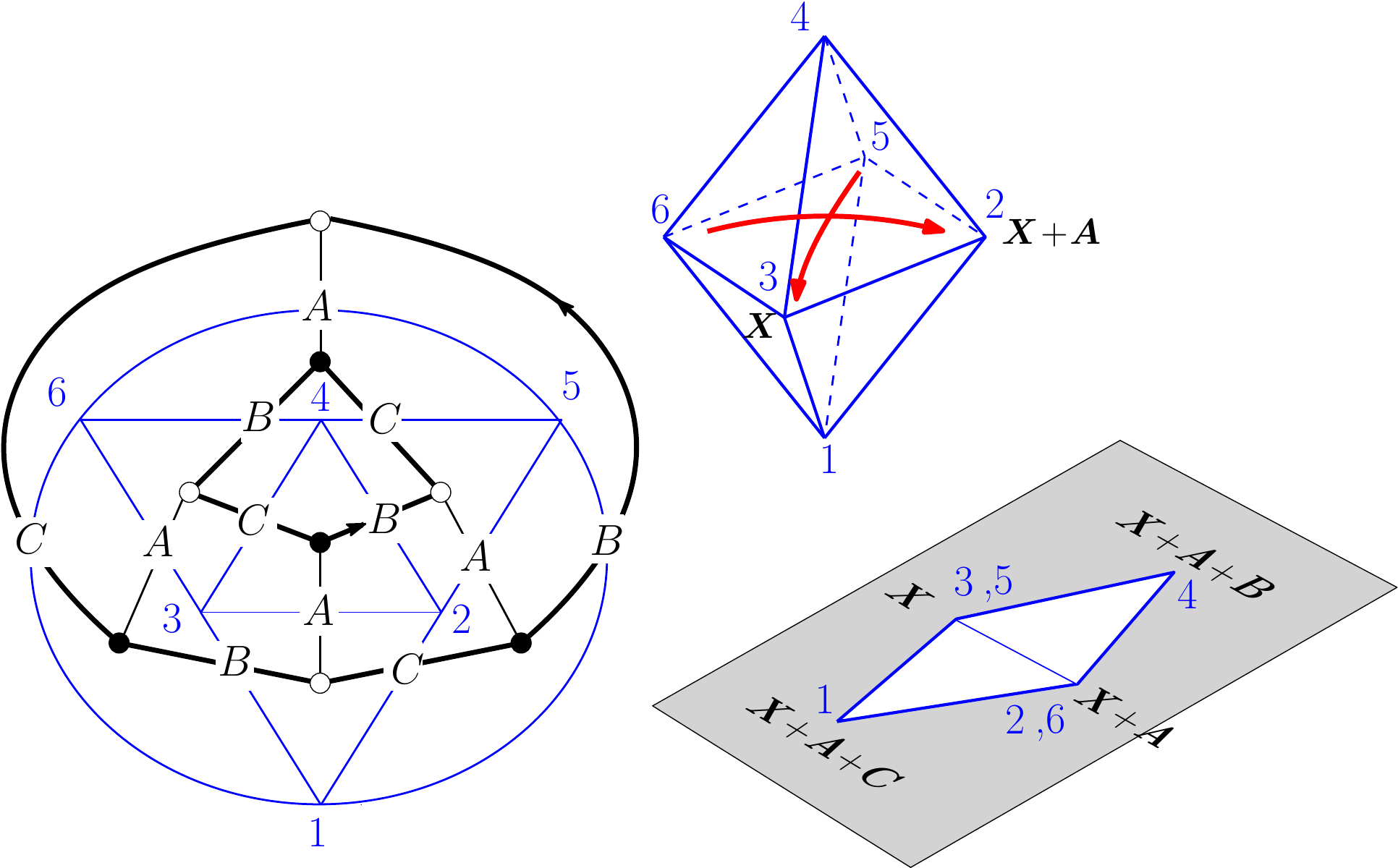}
  \caption{\small An example of bicubic map (black thick lines) with $3$-colored edges and the dual Eulerian triangulation (blue thin lines), 
  with the topology of an octahedron (upper right). The $3$-coloring encodes a folded state of this octahedron in the plane as shown (lower right):
  the 2D height $\X$ associated with the $3$-coloring is nothing but the position in the plane of the vertices of the octahedron after
  folding.}
  \label{fig:folding}
\end{figure}
Planar bicubic maps are dual to planar \emph{Eulerian triangulations}, i.e., maps whose all faces are 
triangles, colored in black and white so that no two adjacent faces have the same color.  A necessary and sufficient
condition for such a coloring to exist is that each vertex of the triangulation has even degree, which is also
the condition for the map to be drawable without lifting the pen, starting and ending at the same vertex. 
This explains the denomination Eulerian. More interestingly, Eulerian triangulations are exactly those
triangulations which, when made of \emph{equilateral triangles} of fixed size (say, with all edge lengths equal to $1/\sqrt{3}$)  
can be embedded into the plane, keeping each triangle equilateral \cite{DFG05}. Any such embedding corresponds to what can be called  
a two-dimensional \emph{folded} state of the triangulation and, in this sense, planar Eulerian triangulations
are exactly those planar triangulations which are foldable in the plane (see Figure~\ref{fig:folding}).

Consider now the FPL$(2)$ model on random planar bicubic maps: as before, it is equivalent to a 3-coloring
of the edges of the map in colors $A$, $B$ and $C$ so that each (trivalent) vertex is incident to edges of different colors.
We can again define up to global translation a 2D height variable $\X$ on each face of the map, 
according to the rules of
Figure~\ref{fig:honeycomb}, and with $\A$, $\B$ and $\C$ as in Figure~\ref{fig:frame}. Moreover, each 3-coloring corresponds to a specific folded state of the dual
Eulerian triangulation and the 2D variable $\X$, attached to the faces of
the bicubic map, hence to the vertices of the triangulation, can be interpreted as the (two-dimensional) 
position of the associated triangulation vertex in the folded state \cite{DFG05}, see Figure~\ref{fig:folding}.    

As in Section~\ref{sec:vortex}, we will consider configurations with magnetic defects $\M$ corresponding
to unvisited vertices (e.g., for $\M=3\A$), or vertices from which several lines emerge (e.g., for $\M=2\A+\B$
or $\M=\A+2\B$). An important difference between the random and regular cases is that we will
not impose that the vertices carrying a defect be trivalent. We will for
instance consider univalent unvisited vertices (defect $\M=\pm \A$) or open paths with
univalent endpoints (defect $\M=\pm \B$ or $\pm \C$), and more generally $m$-valent defects with
arbitrary positive integers $m$ . As a consequence, the set of  magnetic charges $\M$ is no longer restricted 
to the lattice $\mathcal{R}$ but is extended to the larger set $\mathcal{T}=\Z \B+\Z \C$.
Writing as before $\M=\phi_1\A+\phi_2\btwo$, i.e. working in the $(\A,\btwo)$ basis, we may now 
take $\phi_1\in \frac{1}{2}\Z$ (instead of $\frac{3}{2}\Z$), $\phi_2\in \frac{1}{2}\Z$ with still the constraint
that $\phi_1+\phi_2\in\Z$.

\subsection{The KPZ relations}\label{sec:KPZ}
The continuum description of the coupling of 2D quantum gravity to critical matter theories involves
incorporating fluctuations of the underlying metric $\hat{g}$, which is deformed by a multiplicative 
local conformal factor $e^{\gamma_{\scriptscriptstyle{\mathrm{L}}} \varphi_{\scriptscriptstyle{\mathrm{L}}}}$, in terms of a scalar field $\varphi_{\scriptscriptstyle{\mathrm{L}}}$ governed by the Liouville 
action \cite{FD88,DK89}.
For the coupling to gravity of a conformal field theory with central charge $c$, the parameter $\gamma_{\scriptscriptstyle{\mathrm{L}}}$ is fixed 
to the value 
\begin{equation}
\gamma_{\scriptscriptstyle{\mathrm{L}}}=\gamma_{\scriptscriptstyle{\mathrm{L}}}(c)=\frac{1}{\sqrt{6}}\left(\sqrt{25-c}-\sqrt{1-c}\right)\in (0,2] \quad \hbox{for}\quad c\in (-\infty, 1]\ ,
\label{eq:gammaL}
\end{equation}
 by requiring that the (regularized) Liouville random measure $d^2x\, :e^{\gamma_{\scriptscriptstyle{\mathrm{L}}} \varphi_{\scriptscriptstyle{\mathrm{L}}}(x)}:$
is conformally
invariant.  The matter is now subject to the fluctuations of the metric, and in particular matter fields acquire
a multiplicative gravitational dressing of the form $e^{\alpha_{\scriptscriptstyle{\mathrm{M}}}\varphi_{\scriptscriptstyle{\mathrm{L}}}}$ for a suitable value of the ``charge'' $\alpha_{\scriptscriptstyle{\mathrm{M}}}$.

Correlation functions of dressed matter fields are also summed over fluctuations of the metric, hence positions
of the fields are also integrated over in the process. However, correlators still depend on invariants of the
random surfaces generated by the fluctuations of the metric, such as the \emph{area}, given by
$A=\int d^2x \sqrt{|\hat{g}|}  :e^{\gamma_{\scriptscriptstyle{\mathrm{L}}} \varphi_{\scriptscriptstyle{\mathrm{L}}}(x)}:$ and the Einstein action, reduced to the Euler characteristic $\chi=\frac{1}{4\pi} \int d^2x \sqrt{|\hat{g}|}\,
{R}=2-2G$ by the Gauss-Bonnet formula, where $R$ is the scalar curvature and $G$ is the genus of the fluctuating surface. 

In particular, the coupling of a CFT with central charge $c$ to 2D quantum gravity on surfaces 
of fixed genus $G$ and area $A$ results in a new micro-canonical partition function behaving for large $A$ as
$Z_{A,G}\sim \hbox{const. } \mu^A\, A^{(\gamma(c)-2)\chi/2-1}$ in terms of  the ``string susceptibility exponent'' 
$\gamma=1-4/\gamma_{\scriptscriptstyle{\mathrm{L}}}^2$ \cite{KPZ88,FD88,DK89}, namely
\begin{equation}
\gamma=\gamma(c)=\frac{1}{12}\left(c-1-\sqrt{(1-c)(25-c)}\right)\ .
\label{eq:gammac}
\end{equation}
The parameter $\mu=e^\Lambda$ is related to the critical cosmological constant $\Lambda$, the chemical potential 
for the area term in the action.
Note that for planar (genus zero) surfaces, this gives 
\begin{equation}
Z_{A,0}\sim  \hbox{const. }\mu^A\, A^{\gamma(c)-3}\ .
\label{eq:partfunc}
\end{equation}

Likewise, dressed matter conformal primary fields $\Phi_{h,c}$ with classical dimension (or conformal weight)  $h$ acquire a gravitational anomalous dimension
\cite{KPZ88}
\begin{equation}
\Delta(h,c)=\frac{\sqrt{1-c+24\, h}-\sqrt{1-c}}{\sqrt{25-c}-\sqrt{1-c}}\ ,
\label{eq:Deltahc}
\end{equation}
such that non-trivial gravitational $p$-point correlators on random surfaces (of fixed genus) obey the KPZ scaling:
\begin{equation}
\langle \prod_i \Phi_{h_i,c}\rangle_{A} \sim  \hbox{const. } A^{\sum_i \{1-\Delta(h_i,c)\}} \ .
\label{eq:correlator}
\end{equation}

To make contact with our combinatorial problem, we wish to evaluate the large $N$
scaling behavior of various loop models on random planar (bi)cubic maps with $2N$ vertices. 
We therefore set $G=0$. We interpret $2N=A$ as a
measure of the area (i.e., total number of triangles) of the corresponding discretized 
dual (triangulated) random surface. In the case of Hamiltonian cycles (see Figure~\ref{fig:arches}), 
our object of interest has a marked root
edge (or vertex) where we open the loop. The choices of this marking correspond to an overall factor of $2N=A$ 
and we expect therefore a scaling behavior for the partition function of rooted Hamiltonian cycles 
of the form $z_N=A\, Z_{A,0}\sim  \hbox{const. }\mu^A \, A^{\gamma(c)-2}$. As indicated above, the bicubic nature of the graph ensures that
the flat space matter degrees of freedom (the colors $A,B,C$) still describe the gravitational version of the model,
which keeps the {\emph same} central charge $c$ in \eqref{eq:gammac}, with $c=c_{\mathrm{fpl}}(n)$ as in \eqref{eq:cc}.
For $n=0$, we have seen that $c_{\mathrm{fpl}}(0)=-1$ and we recover the asymptotics \eqref{eq:zN} with the exponent
$\gamma=\gamma(-1)$ given by \eqref{eq:gammaval}.

In the following we will compute a number of 2 or 3-point correlators in the discrete model and 
estimate numerically the corresponding scaling behavior (i.e., both the values of $\mu$ 
and of the configuration exponents), which we will compare to the theoretical prediction
\eqref{eq:correlator}, easily rewritten in the present case as
\begin{equation}
Z_{A,0}\,\langle \prod_i \Phi_{h_i,c} \rangle_{A} \sim  \hbox{const. }\mu^A\, A^{\sum_i \{1-\Delta(h_i,c)\} +\gamma(c)-3}\ .
\label{eq:correlatorbis}
\end{equation}

As an illustration of the method, we describe in detail in Appendix~\ref{appendix:cubic} the exact computation of the scaling behavior of 
Hamiltonian cycles on cubic (not necessarily bicolorable) maps, and check the agreement with KPZ scaling at $c=c_{\mathrm{dense}}(0)=-2$.

\subsection{Scaling limits}\label{sec:scalinglimit}
It is widely believed that the \emph{scaling limit} of the critical $\hbox{O}(n)$ model in two dimensions is described by the celebrated \emph{Schramm-Loewner evolution} $\mathrm{SLE}_\kappa$ \cite{OS}, 
and, more precisely, its collection of critical loops by the so-called \emph{conformal loop ensemble} $\mathrm{CLE}_\kappa$ \cite{SS09}. This conformally invariant random process depends on a single parameter $\kappa \geq 0$, 
which in the $\hbox{O}(n)$ model case is  $\kappa=4/g$, with $n=-2\cos (\pi g)$, and $g\in [1/2,1),\,\,\kappa\in (4,8]$ for the \emph{dense} critical phase, and $g\in [1,3/2], \,\,\kappa\in [8/3,4]$ for the \emph{dilute} critical phase \cite{SS09,Duplantier03,MR2112128,zbMATH05541623}. 
(We restrict ourselves here to $n\geq 0$, thus $\kappa\in [8/3, 8]$, the range for which $\mathrm{CLE}_\kappa$ is defined.)
$\mathrm{SLE}_{\kappa}$ paths, which are always non self-crossing, are \emph{simple}, i.e., non-intersecting when $\kappa \in [8/3,4]$, and \emph{non-simple}  
when $\kappa \in (4,8]$ \cite{MR2153402}. The associated $\mathrm{SLE}_\kappa$ central charge is then 
\begin{equation}\label{eq:ckappa}
c=c_{\mathrm{sle}}(\kappa):=\frac{1}{4}(6-\kappa)\left(6-\frac{16}{\kappa}\right) \in [-2, 1]\quad \hbox{for} \quad \kappa \in [8/3, 8]\ .
\end{equation}
This scaling limit has been rigorously established in several  cases: the contour lines of the discrete Gaussian free field, for which $n=2, g=1, \kappa=4$ \cite{10.1007/s11511-009-0034-y}; critical site percolation on the honeycomb lattice \cite{SMIRNOV2001239}, for which $n=1, g=2/3, \kappa=6$; the critical Ising model and its associated Fortuin-Kasteleyn random cluster model on the square lattice \cite{zbMATH05808591,Chelkak2012515} for which,   respectively, $n=1, g=4/3, \kappa =3$ and $n=\sqrt{2}, g=3/4, \kappa=16/3$. 

The fully-packed $\mathrm{FPL}(n)$ model stays in the \emph{same universality class} as the corresponding dense $\hbox{O}(n)$ model, even though its central charge is shifted by one unit as in \eqref{eq:cc} \eqref{eq:cdense}. 
One reason is that the so-called \emph{watermelon} exponents for an \emph{even} number of paths are the same in  $\mathrm{FPL}(n)$ and $\hbox{O}(n)$ models \cite{BN94,BSY94,KdGN96}, and in particular the 2-leg exponent which gives the Hausdorff dimension of the paths.  One is thus led to conjecture that the scaling limit of the fully-packed $\mathrm{FPL}(n)$ model on the honeycomb lattice is described by \emph{space-filling} $\mathrm{SLE}_\kappa$  \cite{zbMATH06814041}, with  $\kappa$ corresponding to the dense $\hbox{O}(n)$ model phase,
\begin{equation}\label{eq:kappan}
\kappa=\frac{4\pi}{\arccos(-n/2)} \in (4,8]\quad \hbox{for} \quad n\in [0,2)\ .
\end{equation}
In the  $\mathrm{FPL}(n=0)$ case, one has $g=1/2, \kappa=8$, so its scaling limit should be given by $\mathrm{SLE}_8$ (or $\mathrm{CLE}_8$), which is a \emph{Peano curve}, that is space-filling.  

Random planar maps, as weighted by the partition functions of critical statistical models, are widely believed to have for scaling limits Liouville quantum gravity (LQG) coupled to the CFT describing these models, or, equivalently, to the corresponding SLE processes. Let us now recall two distinct results associated with the KPZ perspective \cite{KPZ88}.   

The first KPZ relation \eqref{eq:Deltahc} can be rewritten with the help of the Liouville parameter \eqref{eq:gammaL} as the simple quadratic formula,
\begin{equation} \label{eq:KPZdirect}
h(\Delta)=\frac{\gamma_{\scriptscriptstyle{\mathrm{L}}}^2}{4}\Delta^2+\left(1-\frac{\gamma_{\scriptscriptstyle{\mathrm{L}}}^2}{4}\right)\Delta\ .
\end{equation}
Its rigorous proof \cite{springerlink:10.1007/s00222-010-0308-1bis,2009arXiv0901.0277D,rhodes-2008,DRSV12} rests on the sole assumption that the Liouville field $\varphi_{\scriptscriptstyle{\mathrm{L}}}$ and (any)  random fractal curve (possibly described by a CFT) are \emph{independently} sampled. 

The second KPZ result \eqref{eq:gammac} for $\gamma(c)$ \cite{KPZ88}, or equivalently \eqref{eq:gammaL} for $\gamma_{\scriptscriptstyle{\mathrm{L}}}(c)$, gives the precise coupling between the LQG and CFT or SLE parameters. By substituting the SLE central charge $c=c_{\mathrm{sle}}(\kappa)$ of \eqref{eq:ckappa}, one indeed obtains the simple expressions 
\begin{equation} \label{eq:ggLk} 
\gamma=1-\sup\{4/\kappa, \kappa/4\},\quad
\gamma_{\scriptscriptstyle{\mathrm{L}}}=\inf\{\sqrt{\kappa},\sqrt{16/\kappa}\} \ .
\end{equation}
This has been rigorously established in the probabilistic approach by coupling the Gaussian free field in Liouville quantum gravity with SLE martingales \cite{10.1214/15-AOP1055,PhysRevLett.107.131305}.  In the scaling limit, random cluster models on random planar maps can then be shown to converge  (in the so-called peanosphere topology of the mating of trees perspective) to LQG-SLE  \cite{DMS21}. 

This \emph{matching} property \eqref{eq:ggLk} of $\gamma$, $\gamma_{\scriptscriptstyle{\mathrm{L}}}$ and $\kappa$  applies to the scaling limit of the critical, dense or dilute, $\mathrm{O}(n)$ model on a random planar map, as well as  
to the fully-packed $\mathrm{FPL}(n)$ model on random cubic maps. However, on \emph{bicubic} maps, the correspondence \eqref{eq:ggLk} no longer holds, and one then has a \emph{mismatch} \cite{BGS22}, with $c=c_{\mathrm{sle}}(\kappa)$ of \eqref{eq:ckappa}  replaced  in \eqref{eq:gammaL}, \eqref{eq:gammac} and \eqref{eq:Deltahc} by $c=c_{\mathrm{fpl}}(n)=1+c_{\mathrm{sle}}(\kappa)$, with $\kappa$ still given by  \eqref{eq:kappan}. Note that the constraint $c\leq 1$ in the KPZ relations restricts the loop fugacity of the $\mathrm{FPL}(n)$ model on a bicubic map to the range $n\in [0,1]$ with $\kappa \in [6,8]$, while the complementary range $n\in (1,2)$ with $\kappa \in (4,6)$ is likely to correspond to random tree statistics.
 
A coupling between LQG and space-filling SLE with such mismatched parameters has 
yet to be described rigorously. We can simply predict here that for $n\in [0,1]$ the scaling limit of the $\mathrm{FPL}(n)$ model on a bicubic planar map will be given by space-filling $\mathrm{SLE}_\kappa$,   with $\kappa \in [6,8]$ as in \eqref{eq:kappan}, on a $\gamma_{\scriptscriptstyle{\mathrm{L}}}$-LQG sphere with Liouville parameter  
\begin{equation}
\gamma_{\scriptscriptstyle{\mathrm{L}}}=\frac{1}{\sqrt{12}}\left(\sqrt{3\left(\kappa+\frac{16}{\kappa}\right)+22} - \sqrt{3\left(\kappa+\frac{16}{\kappa}\right)-26}\right)\ ,
\label{gammaLbicubic}
\end{equation}
in agreement with conjectures proposed in \cite{BGS22}.

For the $\mathrm{FPL}(n=0)$ model in the bicubic case, we have $\gamma_{\scriptscriptstyle{\mathrm{L}}}=\tfrac{1}{\sqrt{3}}\left (\sqrt{13}-1\right)$, as opposed to $\gamma_{\scriptscriptstyle{\mathrm{L}}}=\sqrt{2}$ in the cubic case, whereas for the bicubic $\mathrm{FPL}(n=1)$ model, to which the next section is devoted, we have $\gamma_{\scriptscriptstyle{\mathrm{L}}}=2$, instead of $\gamma_{\scriptscriptstyle{\mathrm{L}}}=\sqrt{8/3}$ in the cubic case. 

\section{Exponents for the FPL$\boldsymbol{(1)}$ model on bicubic maps}
\label{sec:none}
\begin{figure}
  \centering
  \fig{.40}{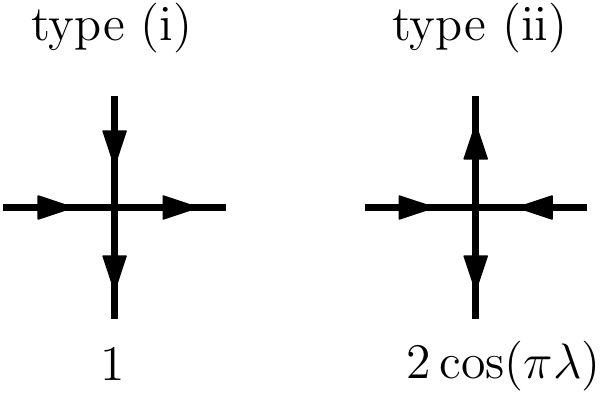}
  \caption{\small The two types of vertex environments for the 6V model on a tetravalent map and the corresponding weights.}
  \label{fig:6Vweights}
\end{figure}

As we shall now see, a direct test of the KPZ formulas for the FPL$(n)$ model on bicubic maps can be performed
in the case $n=1$, whose central charge, given by \eqref{eq:cc} with $g=2/3$, is equal to 
$c=c_{\mathrm{fpl}}(1)=1$. Indeed, as shown in \cite{DFGK99}, the FPL$(1)$ model defined on planar bicubic maps 
is equivalent to a particular instance of the $6$-vertex (6V) model on tetravalent planar maps. 
\begin{figure}
  \centering
  \fig{.85}{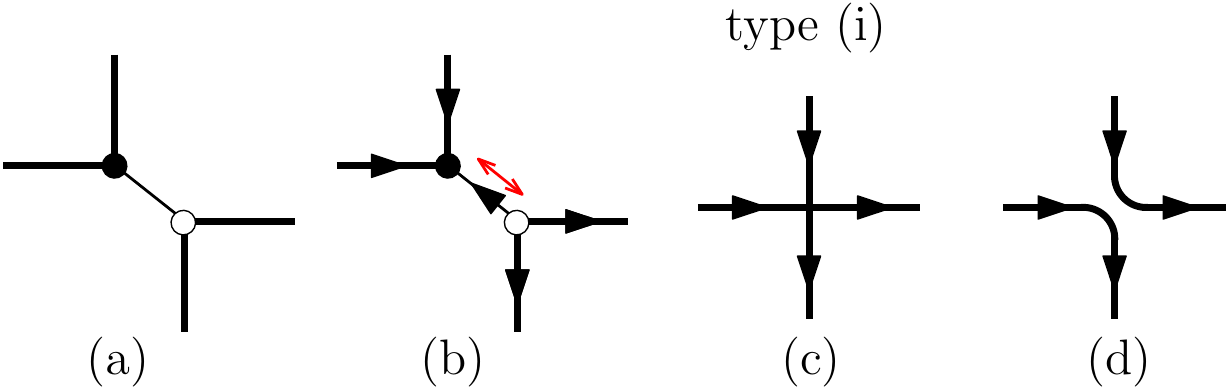}
  \caption{\small In a bicubic map endowed with an unoriented fully packed loop configuration, we consider the pairs
  of black and white vertices linked by an unvisited edge (a). After orienting all incident edges away from the white
  vertex and towards the black one (b), we squeeze the unvisited edge so as to produce a tetravalent vertex of type (i) as
  shown in (c). Doing that for all pairs leads to a 6V configuration on a tetravalent map, where all the vertices are of type (i). The construction is clearly reversible.
  The 6V configuration may itself be transformed into a particular oriented fully packed loop configuration on
  the tetravalent map by untying the type (i) vertices as shown in (d).}
  \label{fig:FPLto6V}
\end{figure}

We recall that the 6V model on tetravalent planar maps consists in orienting each
edge of the map by an arrow, with the so-called \emph{ice rule} that
each vertex of the map is incident to exactly two (half-) edges carrying an incoming arrow
and two (half-) edges carrying an outgoing arrow, see Figure~\ref{fig:6Vweights}.
On a random tetravalent lattice, we may then distinguish between two vertex environments:
for type (i) vertices, the two incoming arrows follow each other when turning around the vertex
while for type (ii) vertices, they are separated by one outgoing arrow
(see Figure~\ref{fig:6Vweights}). Given some fixed $\lambda\in [0,1]$
we attach a weight $2\cos(\pi \lambda)$ to type (ii) vertices 
and $1$ to type (i). In particular, for $\lambda=\frac{1}{2}$, the configurations of arrows
with a non-zero weight are those where all the vertices are of type (i). 
These latter configurations are in bijection with those
of the FPL$(1)$ model on bicubic maps through the following correspondence:
recall that $n=1$ corresponds to the case of fully packed \emph{unoriented} loops without
any attached weight. For any such configuration, let us orient each edge
of the underlying bicubic map from its white extremity to its black one. Note that this
orientation is a property of the bicubic map only and is independent of its loop content.
Now we may squeeze each \emph{unvisited} edge by collapsing its two (black and white) extremities
into a single, uncolored vertex of degree $4$ (see Figure~\ref{fig:FPLto6V}). By doing so, we build a tetravalent map
with oriented edges (corresponding to all the edges originally visited by the loops) and such 
that each vertex is of type (i). The correspondence is one-to-one since we can put back the unvisited edge
by splitting each type (i) vertex into a black and a white connected vertex by pulling the two incident incoming edges 
on one side (defining the black vertex) and the two incident outgoing edges 
on the other side (defining the white vertex), and finally remove all the arrows.

\begin{figure}
  \centering
  \fig{.50}{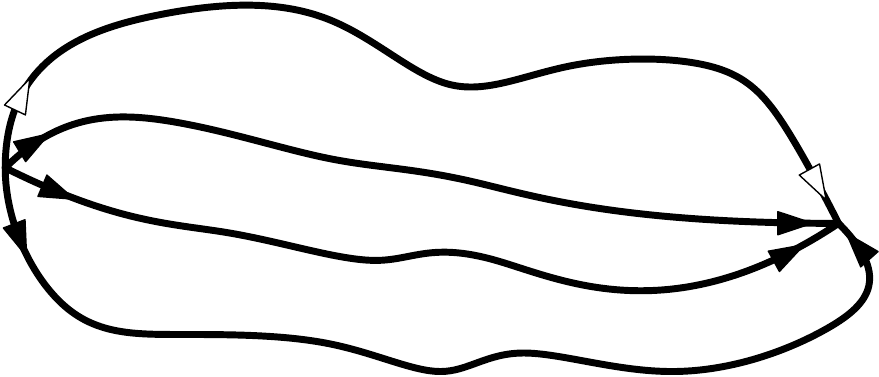}
  \caption{\small Schematic picture of the topology of oriented 6V loops in a watermelon configuration. We have here 
  $\ell=4$ paths, one of which is distinguished (white arrow).}
  \label{fig:watermelon}
\end{figure}

\begin{figure}
  \centering
  \fig{.95}{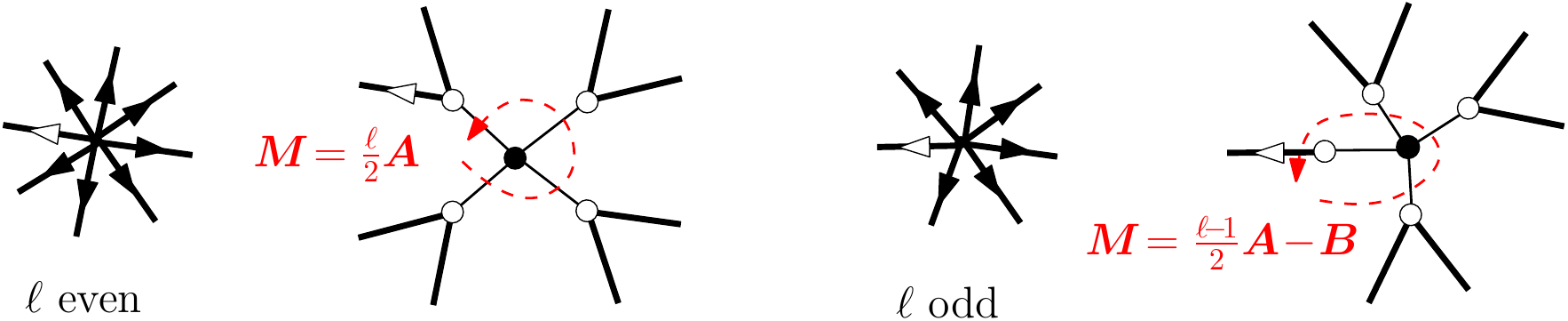}
  \caption{\small Correspondence between a source of $\ell$ lines in the $6V$ model and an unvisited black vertex of degree 
  $\lceil \frac{\ell}{2} \rceil$ in the FPL$(1)$ model.
  The associated magnetic charge $\M$ accounting for the 2D height defect is equal to $\frac{\ell}{2}\A$ for even $\ell$ and to
  $\frac{\ell-1}{2}\A-\B$ for odd $\ell$. }
  \label{fig:6Vdefect}
\end{figure}

The 6V model on tetravalent planar maps with an arbitrary $\lambda\in [0,1]$ was studied in detail by random matrix techniques \cite{KZJ99,K00}
where, as expected, it was shown to correspond to a $c=1$ CFT coupled to gravity. 
 The arrow configurations of the 6V model may themselves be transformed into fully packed oriented loop configurations on the
 underlying tetravalent maps by ``untying'' the vertices (see Figure~\ref{fig:FPLto6V}-(d) in the case of a type (i) vertex).
 Note that these oriented loops are different from the (unoriented) loops of the associated FPL$(1)$ model (these 
 latter loops would correspond instead to paths along which 6V arrow orientations alternate). In the 6V loop language, one may then consider the  \emph{watermelon} configurations, with
 two defects: a source from which $\ell$ oriented lines emerge and 
 a sink at which they all end, see Figure~\ref{fig:watermelon}. The corresponding configuration exponent $\Delta_\ell$ 
 was computed exactly with the result \cite[Eq.~(4.26) with $\lambda=1/2$]{K00}:
 \begin{equation}
 \Delta_\ell=\frac{\ell}{8}\ .
 \label{eq:deltal}
 \end{equation}
Let us now try to recover this result from the general KPZ formulas \eqref{eq:gammac} and \eqref{eq:correlatorbis} in the original FPL$(1)$ language. 
In this language, the source (respectively sink) defects corresponds to an unvisited black (respectively white) vertex
as shown in Figure~\ref{fig:6Vdefect}.  More precisely, for even $\ell$, the unvisited black vertex has degree $\ell/2$ and
thus corresponds to a defect with magnetic charge $\M=\frac{\ell}{2}\A$. For odd $\ell$,  the unvisited black vertex is
connected to $(\ell-1)/2$ regular white trivalent vertices and to a final white bivalent vertex so that the total 
magnetic charge of the defect is now $\M=\frac{\ell-1}{2}\A-\B=\frac{\ell}{2}\A-\frac{1}{2}\btwo$. 
For $n=1$ ($g=\frac{2}{3}$), Equation~\eqref{eq:valexp} becomes 
\begin{equation}
h_{\M}(n=1)=\frac{1}{18}\phi_1^2 +\frac{1}{6}\left(1-\delta_{\phi_2,0}\right)\left(\phi_2^2-\frac{1}{4}\right)\ .
\end{equation}
In particular, we get 
\begin{equation}
\begin{split}
&h_{\frac{\ell}{2}\A}(n=1)=\frac{1}{18}\left(\frac{\ell}{2}\right)^2=\frac{\ell^2}{72}\qquad \hbox{for $\ell$ even}\ ,\\
&h_{\frac{\ell}{2}\A-\frac{1}{2}\btwo}(n=1)=\frac{1}{18}\left(\frac{\ell}{2}\right)^2+\frac{1}{6}\left(\left(-\frac{1}{2}\right)^2-\frac{1}{4}\right)\ =\frac{\ell^2}{72}\qquad  \hbox{for $\ell$ odd}\ ,\\
\end{split}
\end{equation}
hence a result $h_\ell=\frac{\ell^2}{72}$, independently of the parity of $\ell$.

For $c=1$, the KPZ relation \eqref{eq:Deltahc} simplifies into $\Delta(h,1)=\sqrt{h}$. Applying this relation leads us to predict
a configuration exponent for the watermelon configuration equal to:
\begin{equation}
 \Delta(h_\ell,1)=\frac{\ell}{6\sqrt{2}}\ .
 \end{equation}
This value disagrees with the exact value \eqref{eq:deltal}, meaning that the direct application of the KPZ 
formula \emph{does not yield the correct result}.

There is however a simple procedure allowing us to cure the observed discrepancy: let us define  
\begin{equation}
\label{eq:halphadef}
h^{(\alpha)}_{\M}:=\alpha\, \frac{g}{12}\phi_1^2 +\frac{g}{4}\left(1-\delta_{\phi_2,0}\right)\left(\phi_2^2-\left(1-g^{-1}\right)^2\right)\qquad
 \hbox{for}\quad \M=\phi_1 \A+\phi_2\btwo\ , 
\end{equation}
which mimics the expression \eqref{eq:valexp} for $h_{\M}$ by introducing an extra normalization factor $\alpha$ in front
of the first term (i.e., that depending on the component $\phi_1$ in the direction $\A$) with no modification of the 
second term (i.e., that depending on the component $\phi_2$ in the direction $\btwo$).  
We observe immediately that 
\begin{equation}
\begin{split}
&h^{(\alpha)}_{\frac{\ell}{2}\A}(n=1)=\frac{\alpha}{18}\left(\frac{\ell}{2}\right)^2=\frac{\alpha\, \ell^2}{72}\qquad \hbox{for $\ell$ even}\ ,\\
&h^{(\alpha)}_{\frac{\ell}{2}\A-\frac{1}{2}\btwo}(n=1)=\frac{\alpha}{18}\left(\frac{\ell}{2}\right)^2+\frac{1}{6}\left(\left(-\frac{1}{2}\right)^2-\frac{1}{4}\right)\ =\frac{\alpha\, \ell^2}{72}\qquad  \hbox{for $\ell$ odd}\ ,\\
\end{split}
\end{equation}
leading again to the same value $h^{(\alpha)}_\ell=\frac{\alpha\, \ell^2}{72}$ for both parities. In particular, choosing $\alpha=9/8$ leads to
\begin{equation}
\Delta(h^{(9/8)}_\ell,1)=\sqrt{h^{(9/8)}_\ell}=\frac{\ell}{8}=\Delta_\ell\ .
\label{eq:hlneufhuit}
\end{equation}
Otherwise stated, the KPZ relation leads to the correct result provided that we change the exponent $h_{\M}$ into the modified exponent 
$h^{(\alpha)}_{\M}$ with $\alpha=9/8$, where the modification  $h_{\M} \to  h^{(\alpha)}_{\M}$ affects only the part of $h_{\M}$
depending on the component of $\M$ along $\A$ in the $(\A,\btwo)$ basis, with no modification of the part of $h_{\M}$ depending
on the component of $\M$ along $\btwo$. In other words, the compactification radius of the $\phi_1$ component must
be renormalized multiplicatively.

\medskip
This new recipe might seem \emph{ad hoc} but let us make a few comments about it. In the action \eqref{eq:action}, we decided
to attach the \emph{same} ``stiffness'' $g$ to both directions $\psi_1$ and $\psi_2$ of the field $\boldsymbol{\Psi}$
(the $1/3$ factor is only there to correct the fact that $\A$ and $\btwo$ have different norms).  This isotropic choice is natural for $n=2$, 
which describes the pure 3-coloring problem where all colors play the same role but one might question its validity for $n<2$.
A crucial step which led to the expression \eqref{eq:valexpo}, or equivalently \eqref{eq:valexp} for the dimension $h_{\M}$ was 
then the ability to determine the value \eqref{eq:gdef} for this isotropic stiffness $g$. As discussed in \cite{KdGN96}, one
way to fix this value is to demand that the ``electric'' operator with smallest charge in the action \eqref{eq:action} (the $: e^{4\mathrm{i}\pi \psi_2}:$ term) be marginal.
Strictly speaking however, since this criterion involves only the second coordinate $\psi_2$, it only fixes 
the stiffness in the $\psi_2$ direction, leaving that in the $\psi_1$ direction to some undetermined value $g'$ since, as already mentioned, the two directions $\psi_1$ and $\psi_2$ 
are totally independent. On the honeycomb lattice, it seems that $g'=g$ is the correct choice since 
the values $h_M$ \eqref{eq:valexp} obtained by \cite{KdGN96} match with those obtained by Bethe Ansatz methods \cite{BSY94}. It may however occur that, 
when coupled to gravity, the effective stiffness $g'$ takes a different value with a ratio $\alpha:=g'/g\neq 1$ due to metric fluctuations.
If so, this would precisely modify $h_{\M}$ into $h^{(\alpha)}_{\M}$ within the KPZ formula. Verifying this hypothesis 
would require to be able to couple the CG formalism to the fluctuating Liouville field $\varphi_{\scriptscriptstyle{\mathrm{L}}}$  within a unified quantum field theory 
for the three fields $\psi_1$, $\psi_2$ and $\varphi_{\scriptscriptstyle{\mathrm{L}}}$ and to repeat the KPZ arguments in this formalism. 
Still, we present in Section~\ref{sec:discussion} a tentative interpretation of the selected value $\alpha=9/8$ for $n=1$.

\section{Numerics for $\boldsymbol{n=0}$}
\label{sec:numerics}

\subsection{Enumeration methods}
\label{sec:enumeration}
We wish to perform the exact enumeration of Hamiltonian path configurations on planar bicubic maps with a finite number $2N$ of vertices and
with possible magnetic defects. To this end, we use the arch representation displayed in Figure~\ref{fig:arches} or suitable modifications thereof
to account for the desired defects. In all cases, the Hamiltonian path is deformed into a straight line 
with alternating black and white vertices and we must complete it with non-crossing arches above or below the line connecting vertices 
of distinct colors. We used two different ``orthogonal'' enumeration approaches which we describe now.
 
\begin{figure}
  \centering
  \fig{.75}{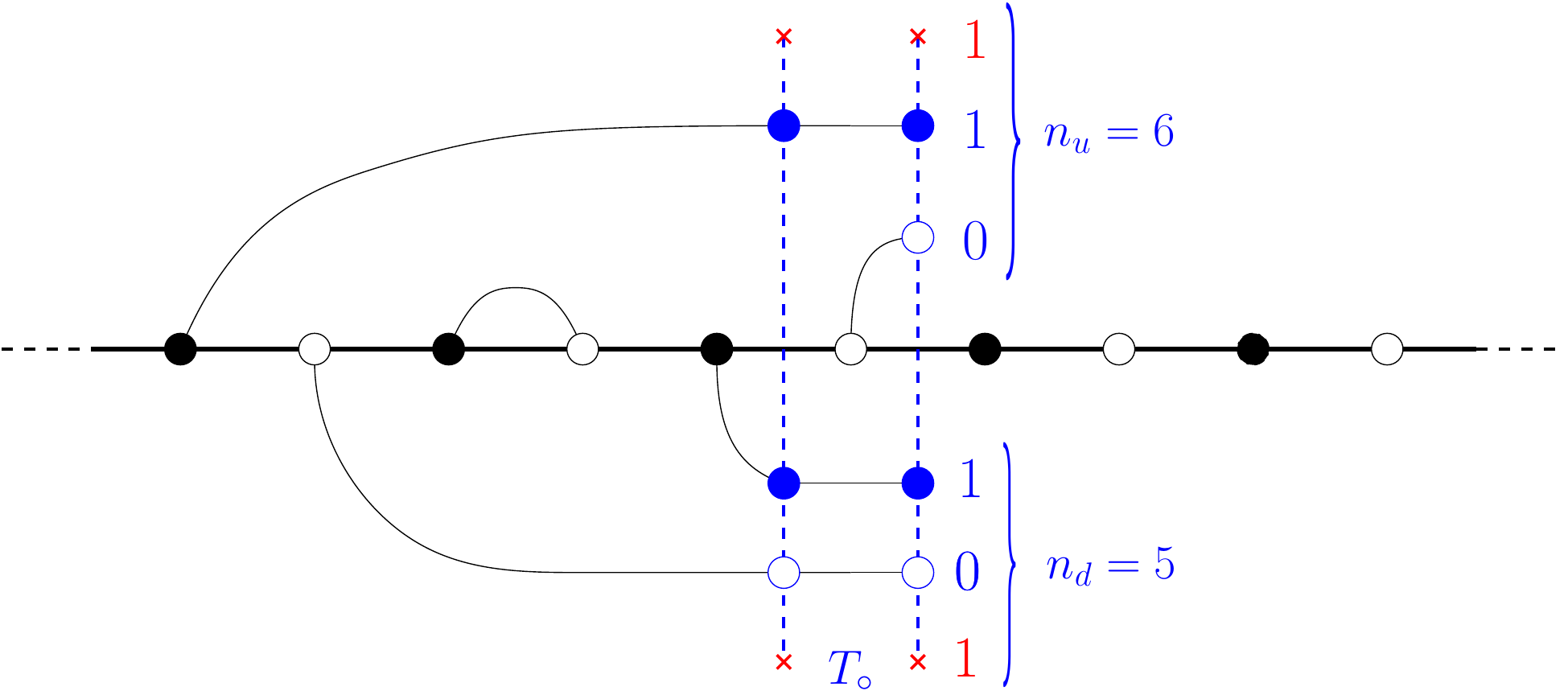}
  \caption{\small Illustration of the transfer matrix method. The arch configuration is built from left to right: each intermediate state (here along
  a blue dashed line) is coded by two positive integers $n_u$ and $n_d$. The passage from one intermediate state to the next corresponds 
  to the action of $T_\circ$ (as shown here) or $T_\bullet$ alternatively along the line. }
  \label{fig:TM}
\end{figure}

\paragraph{The transfer matrix method}
The first approach is a transfer matrix method where we build the arch configurations from left to right along the straight
line of alternating black and white vertices. A configuration is described by the sequence of colors of those arches which have been open but not yet closed: each arch inherits the color of the vertex it originates from,
see Figure~\ref{fig:TM}. We read the upper arch sequence from bottom to top and, if it is made of $p$ arches with 
colors $a_1,\ldots,a_p$ (with $a_i=1$ for black and $0$ for white), we code it via the integer $n_{u}=2^{p}+\sum_{i=1}^p a_i 2^{(i-1)}$.
Similarly, the lower arch sequence read from top to bottom gives a positive integer $n_{d}$, so that the intermediate state may be written
as $|n_{u},n_{d}\rangle$. In this setting, the empty configuration corresponds to the state $|1,1\rangle$ and the number of 
configuration $z_N$ may be written as 
\begin{equation}
z_N=\langle1,1|( T_\circ T_\bullet)^N |1,1\rangle
\end{equation}
with the two elementary transfer matrices $T_\bullet$ and $T_\circ$ defined via
\begin{equation}
\begin{split}
\langle n'_{u},n'_{d}|T_\bullet |n_{u},n_{d}\rangle&=\delta_{n'_{u},2n_{u}+1}\delta_{n'_{d},n_{d}}+
\delta_{n'_{u},n_{u}}\delta_{n'_{d},2n_{d}+1}+\delta_{n'_{u},\frac{n_{u}}{2}}\delta_{n'_{d},n_{d}}+\delta_{n'_{u},n_{u}}\delta_{n'_{d},\frac{n_{d}}{2}}\\
\langle n'_{u},n'_{d}|T_\circ |n_{u},n_{d}\rangle&=\delta_{n'_{u},2n_{u}}\delta_{n'_{d},n_{d}}+
\delta_{n'_{u},n_{u}}\delta_{n'_{d},2n_{d}}+\delta_{n'_{u},\frac{n_{u}-1}{2}}\delta_{n'_{d},n_{d}}+\delta_{n'_{u},n_{u}}\delta_{n'_{d},\frac{n_{d}-1}{2}}\ . \\
\end{split}
\end{equation}
This expression allows us to enumerate $z_N$ up to $N=28$ (see Table~\ref{table:base}) and this approach can be adapted
to situations with (magnetic) defects.

\begin{figure}
  \centering
  \fig{.75}{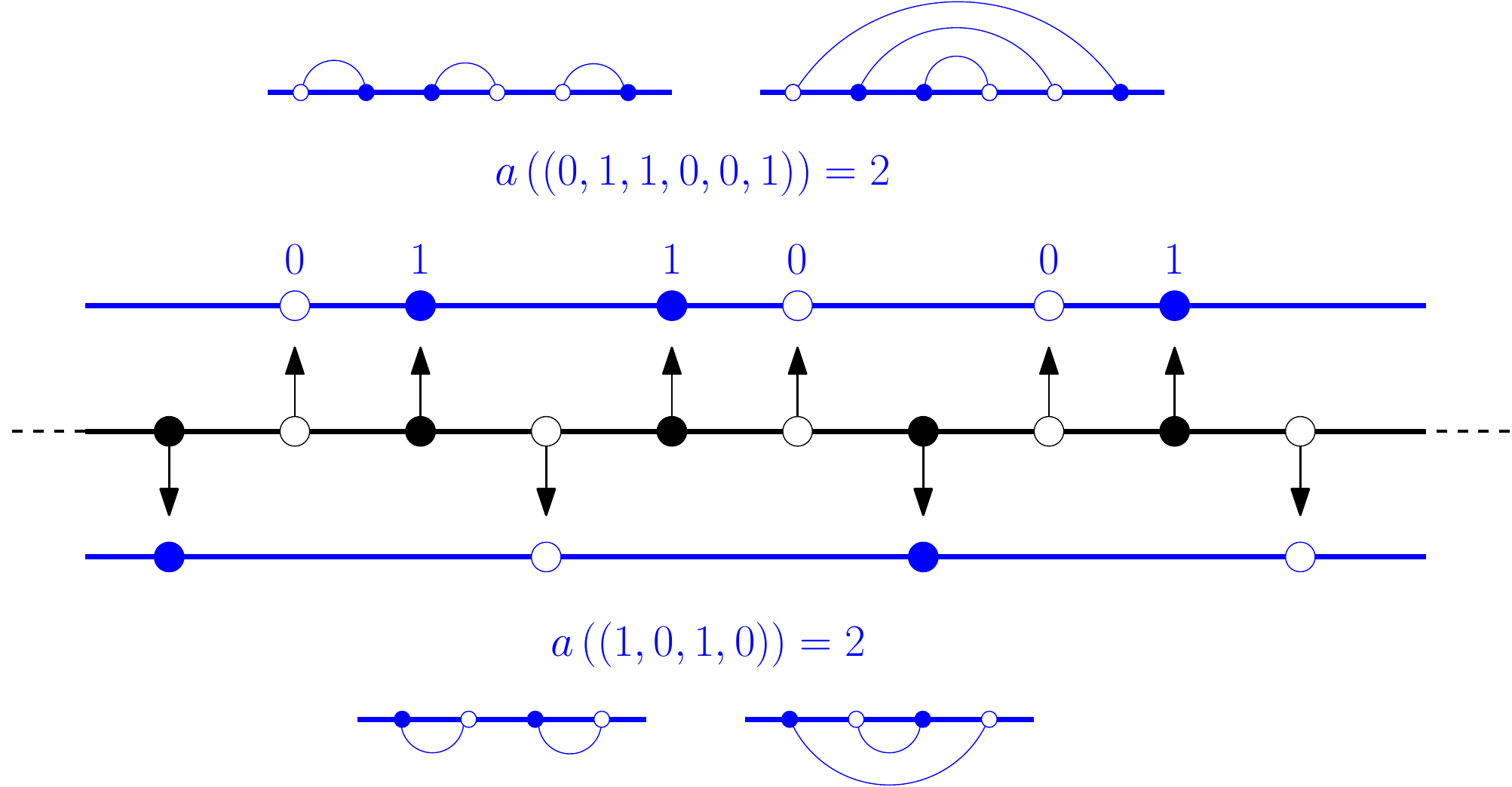}
  \caption{\small Illustration of the up-down factorization method, here for $2N=10$ vertices. We first split the
vertex sequence into complementary sequences $V_{up} = (2,3,5,6,8,9)$ of up-oriented vertices 
and $V_{down} = (1,4,7,10)$, with associated color sequences 
$C_{up}=(0,1,1,0,0,1)$ and $C_{down}=(1,0,1,0)$. There are $a(C_{up}) = 2$ configurations of unisided bicolored 
arches as shown. We also have $a(C_{down}) = 2$ so that there are $ 2 \times 2 = 4$
arch systems compatible with this choice of vertex orientations.
}
  \label{fig:archesUpDown}
\end{figure}

\paragraph{The up-down factorization method} 
 This second approach is closer in spirit to that of \cite{GKN99} and is based on a \emph{two-step} construction
process of Hamiltonian cycles, see Figure~\ref{fig:archesUpDown}.
The first step consists in assigning an up or down orientation to each of the $2N$ vertices
drawn along the straight line.  The second step consists in connecting all the up (resp. down) vertices 
by bicolored (i.e., with endpoints of different colors) arches drawn above (resp. below) the infinite line.
The interest of the method is the following: once the vertex orientations have been fixed, the
system of ``up'' arches and that of ``down'' arches are totally \emph{independent} and the counting
of Hamiltonian cycles is therefore entirely factorized. Moreover, the method is very flexible and it is more easily
adaptable to the case with defects than its transfer matrix counterpart.

More precisely, the $2N$ vertices are numbered by integers $v$ from $1$ to $2N$ according to their position
along the straight line, say from left to right. The color $\hbox{col}(v)$ of the vertex $v$ is nothing
but its parity $\hbox{col}(v) = v \bmod 2$ (with the convention that $\hbox{col}=0$ for white and $\hbox{col}=1$ for black), see Figure~\ref{fig:archesUpDown}. 

In order to fix the up or down orientations of the $2N$ vertices, we split the sequence $(1,2,\ldots,2N)$ 
into two increasing subsequences $V_{up}$ and $V_{down}$. We then read along the line the colors
of the up vertices $C_{up} = \hbox{Col}(V_{up})$ (with, formally, $\hbox{Col}$ the operator acting on a sequence  $V = (v_i)_i$
and returning the sequence $C=\hbox{Col}(V):= (\hbox{col}(v_i))_i$) and do the same for  $C_{down}=\hbox{Col}(V_{down})$.
The number of bicolored arch systems compatible with the choice of orientation $(V_{up}, V_{down})$
is factorized into $a(C_{up}) a(C_{down})$ where $a(C)$ is defined as the number of bicolored
arch systems on \emph{one side} of the line (up and down are clearly equivalent) compatible with the color sequence $C$ (see Figure~\ref{fig:archesUpDown}).

To get the total number of Hamiltonian cycles $z_N$, we have to
sum over all possible partitions $(V_{up}, V_{down})$, hence:
\begin{equation}
  z_N = \sum_{V_{up}} a(\hbox{Col}(V_{up})) a(\hbox{Col}(V_{down}))\ .
 \end{equation}
Non-zero contributions to this sum correspond to \emph{admissible} sequences $V_{up}$ with equal numbers of black and white vertices,
which automatically implies the same property for $V_{down}$.

Denote by $k$ the number of (white/black) vertex pairs in $V_{up}$.
As we have to choose $k$ black vertices among  $N$ and $k$ white vertices
among $N$, the number of admissible partitions to deal with is 
\begin{equation}
     \sum_{k=0}^{N} \binom{N}{k}^2 = \binom{2N}{N}
      \sim \frac{4^N}{\sqrt{\pi N}}\ .
\end{equation}
The computer time therefore increases exponentially
with $N$, which in practice rapidly limits the accessible values of $N$.
One may try to accelerate the program, for instance by using
a ``memoization technique'' which consists in storing the values
of $a(C)$ whenever we encounter the sequence $C$ for the first time
so that we do not have to compute it again at a later occurrence of $C$.
But we are then limited by memory size, since the method requires
the storage of $O(4^N)$ correspondences $(C, a(C))$.

\subsection{Enumeration results}
\label{sec:enumerate}
Let us describe the various configuration ensembles that we have enumerated. In all the definitions below, it is understood that an arch always connects a black and a white vertex and that the straight line is implicitly oriented from left to right.

\paragraph{Hamiltonian cycles}
Our first 
combinatorial quantity is the number $z_N$  of Hamiltonian cycles on planar bicubic maps, as defined earlier, with $2N$ vertices
and with an extra marked visited (root) edge. In the arch language, we may write pictorially 
\begin{equation}
z_N:=\raisebox{-29pt}{\fig{.5}{arches.pdf}}
\end{equation}
with an \emph{infinite line} carrying $2N$ alternating black and white vertices, and with a total of $N$ non-crossing arches.
We have obtained the first values of $z_N$ independently by the transfer matrix and by the up-down factorization methods, 
allowing for a cross-check of the results. We find:
\begin{equation}
(z_N)_{N\geq 1}=(2,8,40,228,1424,\ldots )\ .
\end{equation}
The complete list up to $N=28$ is given in Table~\ref{table:base} of Appendix~\ref{appendix:results}. It confirms and extends the results of \cite{GKN99} (limited to $N\leq 20$), 
see also the sequence \href{http://oeis.org/A116456}{A116456} in OEIS~\cite{OEIS}.

\paragraph{Hamiltonian open paths with trivalent endpoints}
The second quantity that we considered is the number $y_N$ of Hamiltonian \emph{open} paths on planar bicubic maps with $2N+2$ vertices
(in particular, the endpoints of the path are trivalent). As already seen 
in the regular lattice case, the endpoints of the path correspond to defects with opposite magnetic charges $\pm \M$, with $\M=2\A+\B=\frac{3}{2}\A+\frac{1}{2}\btwo$.
In the arch language, we may write pictorially 
\begin{equation}
y_N:=\raisebox{-24pt}{\fig{.52}{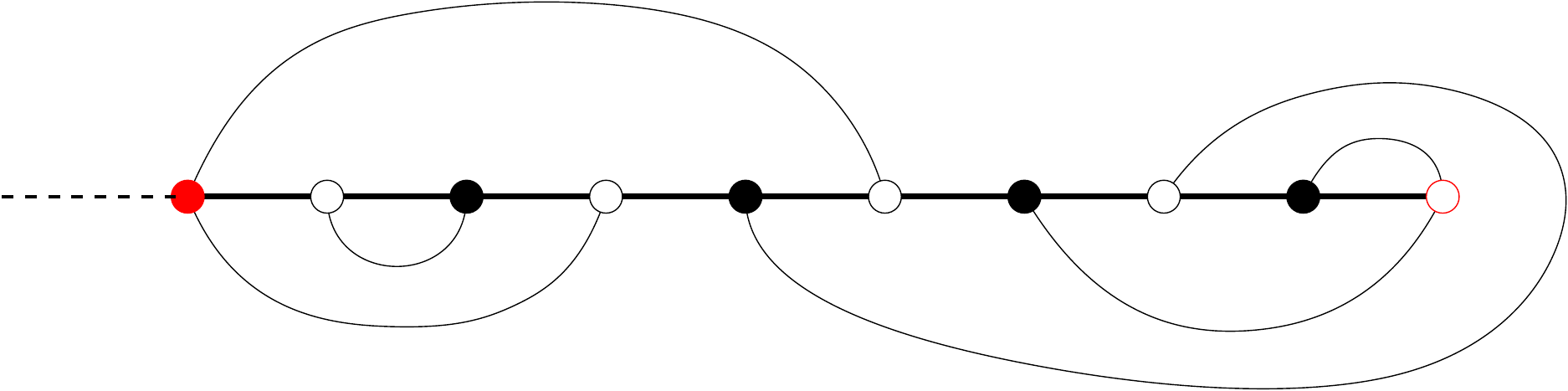}}
\end{equation}
with a now a \emph{line segment} of $2N+2$ alternating black and white vertices and a total of $N+2$ non-crossing arches. 
In the planar representation of the map, we fixed as \emph{external face} that containing the corner between the two unvisited edges at the black endpoint
(marked here by a dashed segment -- this de facto allows us to extend the line segment into an infinite half-line). The arches are now allowed to wind around the line segment by passing to the right of the white endpoint.
We obtained the first values of $y_N$ independently by the transfer matrix and by the up-down factorization methods:
\begin{equation}
(y_N)_{N\geq 0}=(1,6,40,286,2152,\ldots )\ .
\end{equation}
The complete list up to $N=16$ is given in Table~\ref{table:semi-trivalent} of Appendix~\ref{appendix:results}. 

\paragraph{Hamiltonian open paths with univalent endpoints}
The third quantity of interest is the number $x_N$ of Hamiltonian open paths on planar bicolored maps with $2N$ trivalent vertices and
$2$ univalent ones. There the path necessarily starts and ends at the two univalent vertices which moreover correspond to defects with opposite 
magnetic charges $\pm \M$ with now $\M=\B=-\frac{1}{2}\A+\frac{1}{2}\btwo$.
In the arch language, we have pictorially 
\begin{equation}
x_N:=\raisebox{-22pt}{\fig{.52}{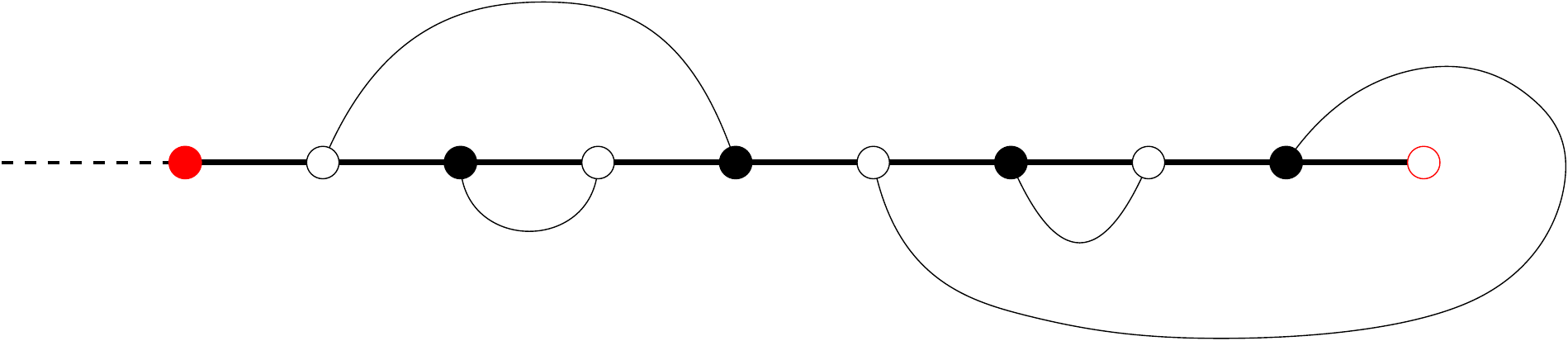}}
\end{equation}
with a line segment of $2N+2$ alternating black and white vertices and a total of $N$ non-crossing arches
connecting all vertices except the two extremal ones. In the planar representation of the map, we took as external face that containing the corner
at the univalent black vertex so that the arches are allowed to wind around the line segment by passing to the right of the white univalent vertex.
We obtained the first values of $x_N$ independently by the transfer matrix and by the up-down factorization methods:
\begin{equation}
(x_N)_{N\geq 0}=(1,4,24,168,1280,\ldots )\ .
\end{equation}
The complete list up to $N=17$ is given in Table~\ref{table:semi} of Appendix~\ref{appendix:results}.

\bigskip

The following ensembles correspond to ``vacancy defects'', i.e.~\ configurations having two or three \emph{unvisited vertices}.
\paragraph{Hamiltonian cycles with two unvisited univalent vertices}

The fourth quantity which we studied is the number $2 w_N$ of cycles on planar bicolored maps with $2N$ trivalent vertices and
$2$ univalent ones, such that the cycle visits all the trivalent vertices but, since it is a cycle, cannot visit the univalent ones (which are then necessarily of different
colors). This situation 
corresponds to having two defects with opposite magnetic charges $\pm \M$ with $\M=\A$.
In the arch language, we have pictorially 
\begin{equation}
w_N:=\frac{1}{2}\times\raisebox{-11pt}{\fig{.52}{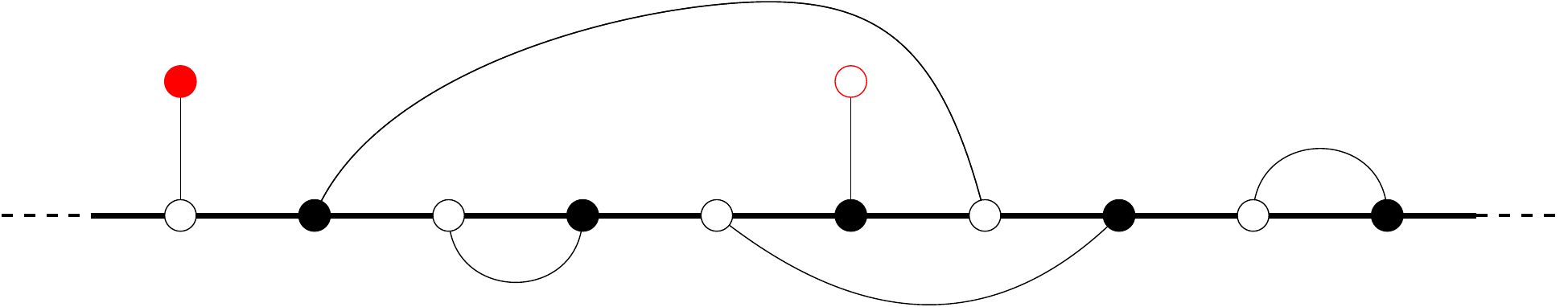}}
\end{equation}
with an infinite line carrying $2N$ alternating white and black vertices, a black univalent vertex grafted above the first 
(white) vertex of the line and a white univalent vertex grafted above or below one of the black vertices of the line. 
The configuration now has a total of $N-1$ non-crossing arches.
In the planar representation of the map, we took as external face that containing the corner
at the univalent black vertex. The factor $\frac{1}{2}$ is because we factored out the trivial symmetry consisting in flipping
up or down the univalent white vertex. 
The first values of $w_N$ were obtained independently by the transfer matrix method and by the up-down factorization method, with
result:
\begin{equation}
(w_N)_{N\geq 1}=(1,4,22,140,972,\ldots )\ .
\end{equation}
The complete list up to $N=18$ is given in Table~\ref{table:m+1m-1} of Appendix~\ref{appendix:results}.

\paragraph{Hamiltonian cycles with two unvisited bivalent vertices}
Our fifth quantity is the number $v_N$ of cycles on planar bicolored maps with $2N$ trivalent vertices and
$2$ bivalent ones (which are then necessarily of different
colors), where we require that the cycle visits all the trivalent vertices but \emph{not the bivalent ones}. This situation 
corresponds to having two defects with opposite magnetic charges $\pm \M$ with $\M=2\A$.
In the arch language, we have pictorially 
\begin{equation}
v_N:=\raisebox{-19pt}{\fig{.52}{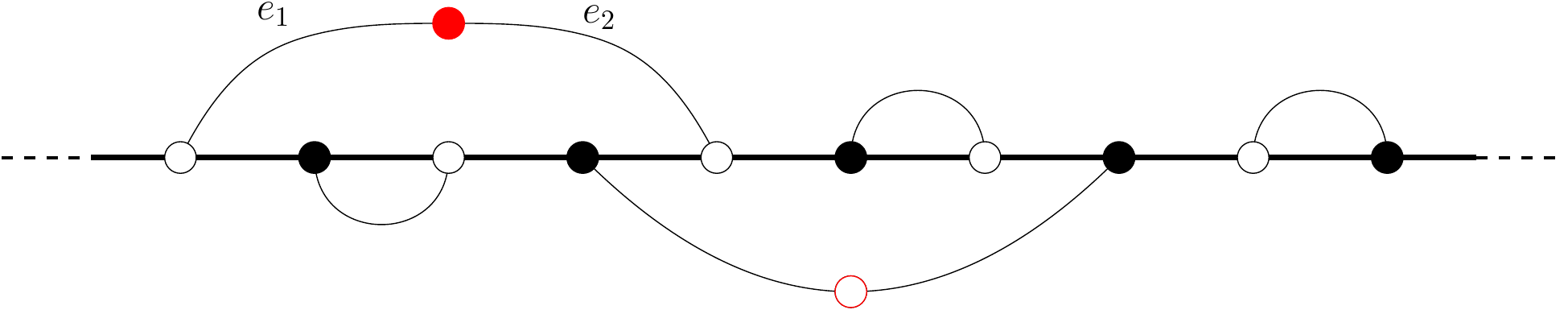}}
\end{equation}
with an infinite line carrying $2N$ alternating white and black vertices, a black bivalent vertex linked (from above)
to two (white) vertices of the line among which we choose the first white vertex along the line, and a white bivalent vertex linked to two black 
vertices of the line. 
The configuration has $N-2$ additional non-crossing arches. The two edges incident to the bivalent black vertex are distinguished as $e_1$ and $e_2$
and, in the planar representation of the map, we take as external face that containing the corner between edges $e_1$ and $e_2$ clockwise. 
The first values of $v_N$ were obtained by the up-down factorization method, with result:
\begin{equation}
(v_N)_{N\geq 2}=(1,10,84,682,5534,\ldots )\ .
\end{equation}
The complete list up to $N=21$ is given in Table~\ref{table:m+2m-2} of Appendix~\ref{appendix:results}.
\paragraph{Hamiltonian cycles with two univalent one bivalent unvisited vertices}
The last quantity which we considered is the number $4u_N$ of cycles on planar bicolored maps with $2N$ trivalent vertices, 
$1$ bivalent black one and $2$ univalent white ones, such that the cycle visits all the trivalent vertices but not the bivalent or
univalent ones. This situation 
corresponds to having \emph{three defects} with respective magnetic charges $2\A$, $-\A$ and $-\A$.
In the arch language, we have pictorially 
\begin{equation}
u_N:=\frac{1}{4}\times \raisebox{-19.pt}{\fig{.52}{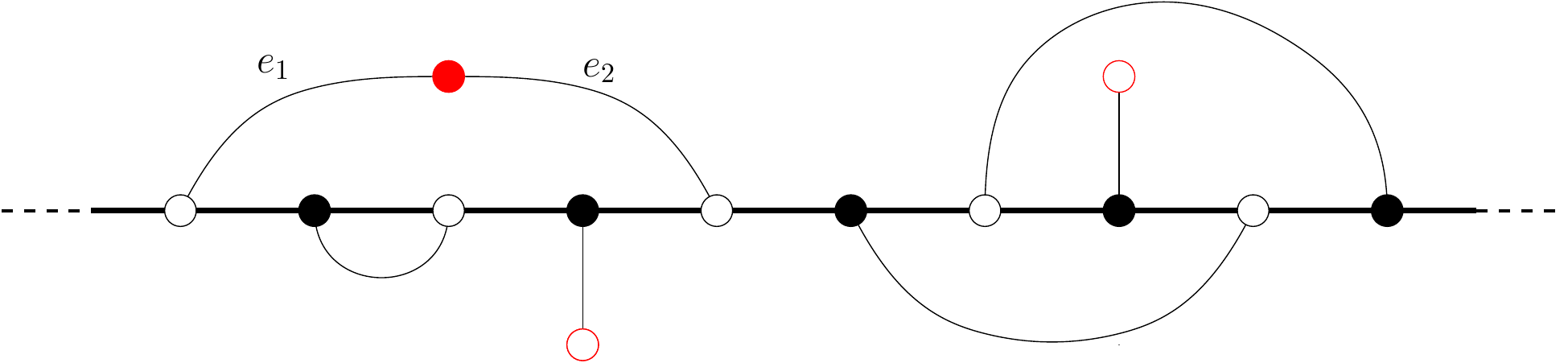}}
\end{equation}
with an infinite line carrying $2N$ alternating white and black vertices, a black bivalent vertex linked (from above)
to two (white) vertices of the line among which we choose the first white vertex along the line, and two white univalent vertices grafted to black 
vertices along the line. 
The configuration has $N-2$ additional non-crossing arches. Again, the two edges incident to the bivalent black vertex are distinguished as 
$e_1$ and $e_2$
and, in the planar representation of the map, we take as external face that containing the corner between edges $e_1$ and $e_2$ clockwise.
The factor $\frac{1}{4}$ is because we factored out the trivial symmetry consisting in flipping
up or down the univalent white vertices. 
The first values of $u_N$ were obtained by the up-down factorization method, with result:
\begin{equation}
(u_N)_{N\geq 2}=(1,10,90,798,7094,\ldots )\ .
\end{equation}
The complete list up to $N=17$ is given in Table~\ref{table:m+2m-1m-1} of Appendix~\ref{appendix:results}.

\subsection{Exponential growth rate}
\label{sec:growthrate}
All the quantities $t_N=z_N,y_N,x_N,w_N,v_N,u_N$ which we introduced so far are expected to have the asymptotic behavior
 \begin{equation}
 t_N\sim \hbox{const.}\ \frac{\mu^{2N}}{N^{\beta_t}}\ ,
 \label{eq:tN}
 \end{equation}
 with \emph{the same} exponential growth rate $\mu$ and sub-leading corrections characterized by an exponent
 $\beta_t$ \emph{specific to each quantity} and whose value should be predicted from the KPZ formulas.
 In order to evaluate $\mu$ and $\beta_t$, we construct from the sequence $t_N$ the following two sequences
 \begin{equation}
 a_{N}:=\frac{t_{N+1}}{t_N}\ , \qquad b_{N}:=N^2\, \hbox{Log}\frac{t_{N+2}t_N}{(t_{N+1})^2}\ ,
 \end{equation}
which are such that 
\begin{equation}
a_{N}\underset{N\to\infty}{\rightarrow} \mu^2\ , 
\qquad b_{N}\underset{N\to\infty}{\rightarrow} \beta_t\ . 
\end{equation}
To improve our estimates, we have recourse to two \emph{series acceleration methods}, used for accelerating the rate of convergence of our 
sequences above. Both are expressed via the iterated finite difference operators $\Delta^k$, $k\in \N^*$, defined by
\begin{equation}
\begin{split}
&(\Delta f)_N:=f_{N+1}-f_N\ ,\\
&(\Delta^2 f)_N:=(\Delta(\Delta f))_N=f_{N+2}-2f_{N+1}+f_N\ , \ \ldots  \\ 
\end{split}
\end{equation}
The first method considers the sequences
\begin{equation}
\tilde{a}^{(k)}_N:=\frac {1}{k!}\left(\Delta^k \hat{a}^{(k)}\right)_N \quad \hbox{with} \quad \hat{a}^{(k)}_N:=N^k a_N
\label{eq:deftildea}
\end{equation}
and similarly defined sequences $\tilde{b}^{(k)}$. The convergence of these sequences to $\mu^2$ and $\beta_t$ is faster
for increasing $k$ even though, in practice, since we know $t_N$ for the first values of $N$ only, we cannot go to $k$ larger that $7$ or so.

The second method is the so-called Aitken-$\Delta^2$ method which considers sequences defined recursively via
\begin{equation}
\bar{a}^{(k)}_N:=\bar{a}^{(k-1)}_N-\frac {k+1}{k}\frac{\left(\Delta \bar{a}^{(k-1)}\right)_N\left(\Delta \bar{a}^{(k-1)}\right)_{N-1}}{\left(\Delta^2 \bar{a}^{(k-1)}\right)_{N-1}}\quad \hbox{with} \quad \bar{a}^{(0)}_N:=a_N
\label{eq:defbara}
\end{equation}
and similarly defined sequences $\bar{b}^{(k)}$. Again, the convergence of these sequences is faster
for increasing $k$ (in practice we use $k=1,2$ and $3$).

\begin{figure}[h]
  \centering
  \fig{.75}{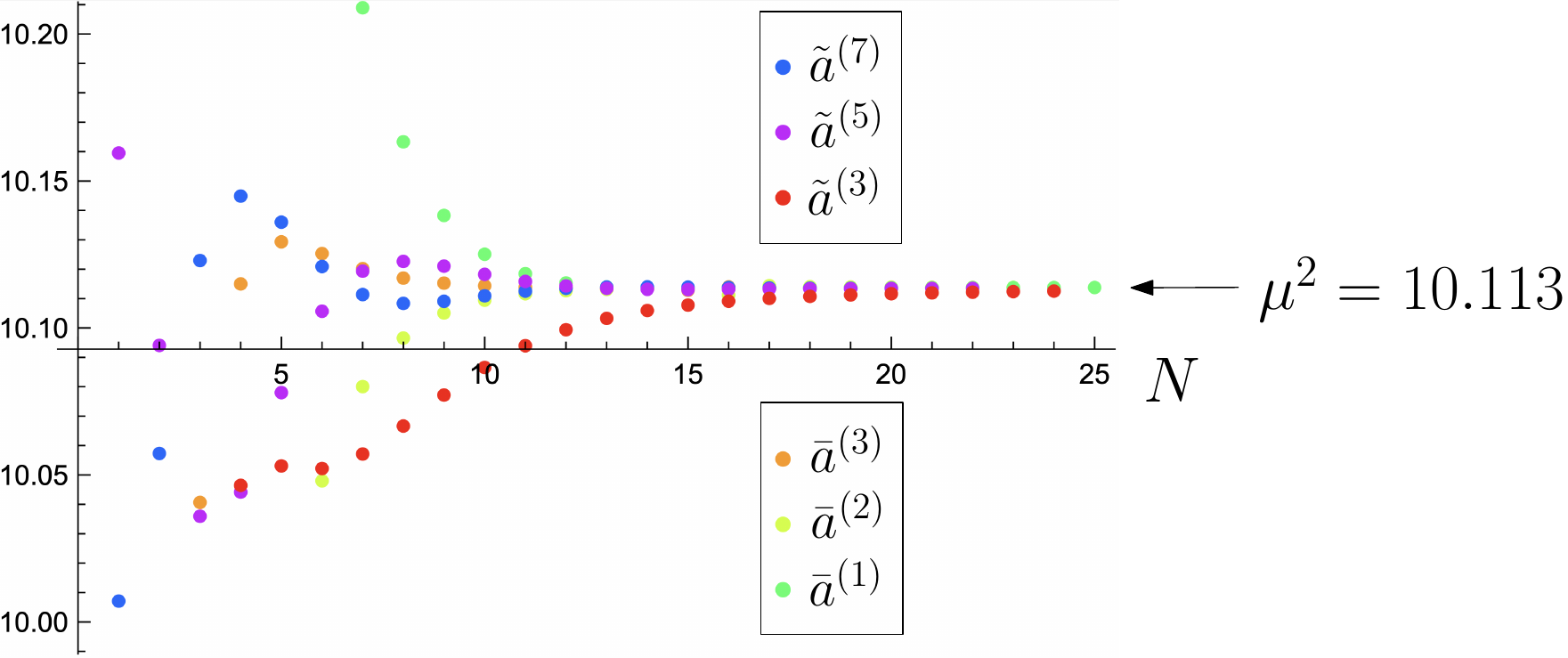}
  \caption{\small Estimate of $\mu^2$ from the sequences $\tilde{a}^{(k)}$ for $k=3,5,7$ and the sequences $\bar{a}^{(k)}$ for $k=1,2,3$
  as defined via  \eqref{eq:deftildea} and \eqref{eq:defbara} with $a_N=z_{N+1}/z_N$.}
  \label{fig:mufromz}
\end{figure}

\paragraph{Estimate of $\boldsymbol{\mu}$ from the sequence $\boldsymbol{z_N}$.}
Figure~\ref{fig:mufromz} shows our estimates for $\mu^2$ obtained from our data $z_N$ for the numbers of Hamiltonian cycles. 
More precisely, it displays the sequences $\tilde{a}^{(k)}$ for $k=3,5,7$ and $\bar{a}^{(k)}$ for $k=1,2,3$ using the sequence
$a_N=z_{N+1}/z_N$ as original input. We estimate from these data
\begin{equation}
\mu^2=10.113\pm0.001\ ,
\end{equation} 
so that $\hbox{Log}(\mu^2)=2.3138\pm 0.0001$, in agreement with \cite{GKN99}.

\begin{figure}[h]
  \centering
  \fig{.75}{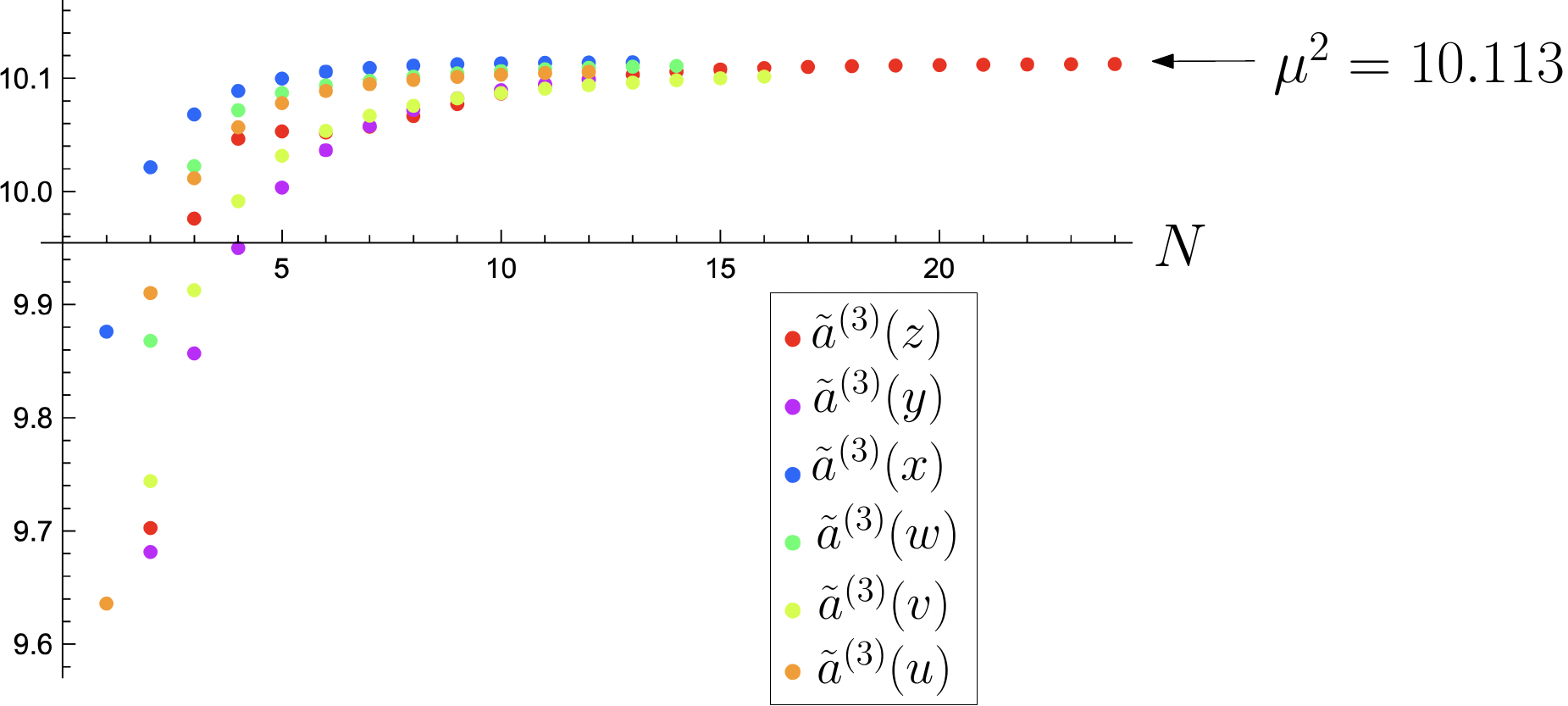}
  \caption{\small Estimate of $\mu^2$ from the sequences $\tilde{a}^{(3)}(t)$ 
  as defined via  \eqref{eq:deftildea} for $a_N=t_{N+1}/t_N$ and $t_N=z_N,y_N,\ldots,u_N$.}
  \label{fig:mucompar}
\end{figure}

\paragraph{Estimate of $\boldsymbol{\mu}$ from other observables.}  We then compared this value of $\mu$ with that obtained from the
other sequences $y_N,x_N,\ldots,u_N$ corresponding to the enumerations of the various Hamiltonian path configurations with defects
introduced in the previous section.  Figure~\ref{fig:mucompar} shows the values of $\mu^2$ obtained from the sequences $\tilde{a}^{(3)}(t)$ 
obtained via \eqref{eq:deftildea} for $a_N=t_{N+1}/t_N$ with $t_N=y_N,x_N,\ldots,u_N$ and compare them with that obtained previously 
for $t_N=z_N$. 
As expected, all estimates converge to the \emph{same value} of $\mu$, which is a non-trivial test of the consistency of our data for
the various sequences.

\subsection{Exponents}
\label{sec:exponents}
\begin{figure}[h]
  \centering
  \fig{.75}{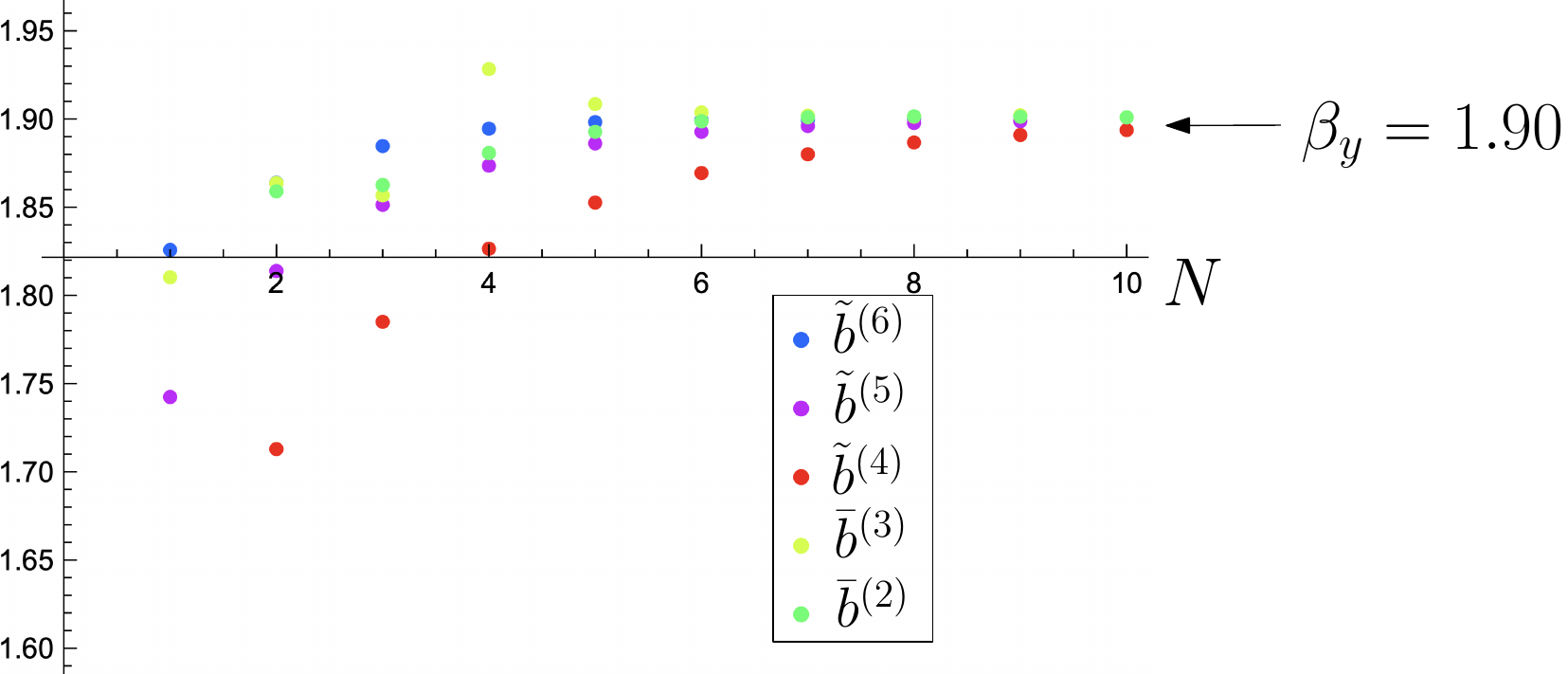}
  \caption{\small Estimate of $\beta_y$ from the sequences $\tilde{b}^{(k)}$ for $k=4,5,6$ and the sequences $\bar{b}^{(k)}$ for $k=2,3$
with $b_N=N^2\, \hbox{Log}(y_{N+2}y_N/(y_{N+1})^2)$.}
  \label{fig:betayest}
\end{figure}

Let us now present our numerical estimates for the various exponents $\beta_z$, $\beta_y $, $\beta_x$, $\beta_w$, $\beta_v$ and $\beta_u$ 
as defined by \eqref{eq:tN} with $t_N=z_N$, $y_N$,  $x_N$,  $w_N$,  $v_N$ and $u_N$ respectively. Figure~\ref{fig:betayest} shows
for instance the estimate $\beta_y=1.90\pm0.01$ deduced from the sequences $\tilde{b}^{(k)}$ for $k=4,5,6$ and $\bar{b}^{(k)}$ for $k=2,3$
with $b_N=N^2\, \hbox{Log}(y_{N+2}y_N/(y_{N+1})^2)$ as input. 
Repeating this analysis with our numerical data for the various sequences leads to
\begin{equation}
\begin{split}
& \beta_z=2.77\pm0.01 \ ,\qquad \beta_y=1.90\pm0.01 \ ,\qquad \beta_x= 1.19\pm0.01\ ,\\
& \beta_w=1.99\pm0.01 \ ,\qquad \beta_v=2.38 \pm0.03 \ ,\qquad \beta_u= 1.32\pm0.02\ .
\label{eq:numestimates}
\end{split}
\end{equation}
Here the announced values correspond to the \emph{stable} digits in $(\tilde{b}^{(k)})_{N_{max}^{(k)}}$ for the largest
accessible value of $N_{max}^{(k)}$ at a given $k$, i.e., those digits which do not vary when going from $N_{max}^{(k)}-1$ to $N_{max}^{(k)}$
nor by increasing $k$ by $1$. This implicitly assumes that we have already reached the asymptotic regime for $N\sim N_{max}^{(k)}$.
From our data, this condition is not satisfied for $\beta_v$ and its value announced above is probably slightly underestimated. We will
come back to this point in the next section.
As for the indicated error ranges, they are estimated from the observed amplitude for the variation of the first digit which is not yet stabilized.

\section{Comparison with KPZ predictions}
\label{sec:comparison}
We now wish to compare the numerical exponents above to their values predicted by the KPZ equivalence.
Since we expect that $c=c_{\mathrm{fpl}}(0)=-1$ for $n=0$ ($g=\frac{1}{2}$), we have, from \eqref{eq:gammac}, the exponent
\begin{equation}
\gamma:=\gamma(-1)=-\frac{1+\sqrt{13}}{6}= -0.76759\ldots
\label{eq:KPZgamma}
\end{equation}
and, from \eqref{eq:Deltahc}, the gravitational anomalous dimensions
\begin{equation}
\Delta_{\M}:=\Delta(h_{\M},-1)=\frac{\sqrt{1+12 h_{\M}}-1}{\sqrt{13}-1}\ ,
\label{eq:Delta0h}
\end{equation}
with, from \eqref{eq:valexp} at $n=0$ ($g=\frac{1}{2}$),
\begin{equation}
h_{\M}=\frac{1}{24}\phi_1^2 +\frac{1}{8}\left(1-\delta_{\phi_2,0}\right)(\phi_2^2-1)
\qquad \hbox{for}\quad \M=\phi_1 \A+\phi_2\btwo \ .
\label{eq:hM0}
\end{equation}
From the general formulas \eqref{eq:partfunc} and \eqref{eq:correlatorbis}
and the identity $h_{-\M}=h_{\M}$, hence $\Delta_{-\M}=\Delta_{\M}$, we may write
\begin{equation}
\begin{matrix}
\label{eq:KPZpred}
\beta_z=2-\gamma\ , \quad \hfill &\beta_y=1+2\Delta_{\frac{3}{2}\A+\frac{1}{2}\btwo}-\gamma\ , \quad \hfill&
\beta_x= 1+2\Delta_{-\frac{1}{2}\A+\frac{1}{2}\btwo}-\gamma\ , \hfill  \\ \beta_w=1+2\Delta_{\A}-\gamma 
\ , \quad \hfill &\beta_v= 1+2\Delta_{2\A}-\gamma\ , \quad \hfill  &\beta_u= \Delta_{2\A}+2\Delta_{\A}-\gamma\ . \hfill 
\end{matrix}
\end{equation}

\medskip
As in \cite{GKN99}, our estimated value $\beta_z=2.77\pm0.01$ above is in perfect agreement 
with the announced value $2-\gamma=2.76759\ldots$ and we thus confirm the prediction \eqref{eq:gammaval} of \cite{GKN99}.

\begin{figure}[h]
  \centering
  \fig{.55}{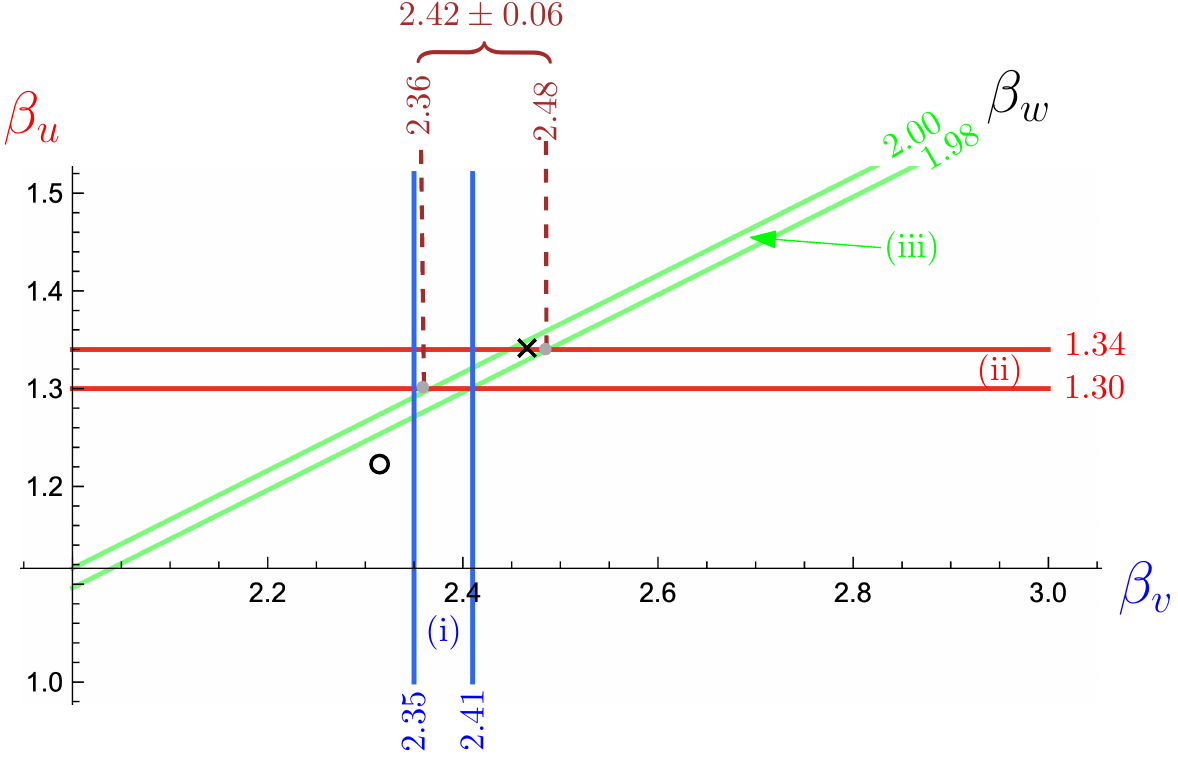}
  \caption{\small Consistency between the estimated value of  $\beta_v$ (domain (i) between the two vertical blue lines),
 that of $\beta_u$ (domain (ii) between the two horizontal red lines) and that of the combination $2\beta_u-\beta_v$ 
 (domain (iii) between the two diagonal lines) as obtained from the relation \eqref{eq:relbeta} and the estimated range 
 \eqref{eq:numestimates} for $\beta_w$. The three domains do indeed share a common sector in the $(\beta_v,\beta_u)$ plane.
 In the absence of a direct estimate (i) for $\beta_u$, we would predict from the intersection of domains (ii) and (iii) only a 
 somewhat larger domain $\beta_v=2.42\pm 0.06$ (dashed brown lines). The $\circ$ symbol indicates the
 prediction of the naive KPZ formulas while the \raisebox{1.pt}{$\scriptstyle{\times}$} symbol indicates that of the $(4/3)$-corrected KPZ formulas,
 see text. }
 \label{fig:Betacheck}
\end{figure}

\bigskip
Before going to the precise values of the other exponents, we note that we have from \eqref{eq:KPZpred} the consistency relation
\begin{equation}
2\beta_u-\beta_v=2\beta_w+\gamma-3\ .
\label{eq:relbeta}
\end{equation}
Assuming now that $\gamma$ is indeed determined by \eqref{eq:KPZgamma}, this relation provides a cross check
between, on the one hand, our numerical estimates for $\beta_u$ and $\beta_v$ and, on the other hand, the estimate
for $2\beta_u-\beta_v$ inherited from the estimate
 \eqref{eq:numestimates} for $\beta_w$. As displayed in Figure~\ref{fig:Betacheck}, these three estimates are indeed consistent with each other.
 As mentioned above, we are however not fully confident with our estimated range of $\beta_v$. The relation \eqref{eq:relbeta} may then be used to 
determine the value of $\beta_v$ from those of $\beta_u$ and $\beta_w$. If we do so,
this leads us to reevaluate the estimate for $\beta_v$ as $\beta_v=2.42\pm0.06$, see Figure \ref{fig:Betacheck}. 
 
\begin{table} [h]
    \centering
       \begin{tabular}{|c|c|c|c|}
        \hline
  \null   &         numerics  &   naive KPZ ($\alpha=1$)   &   $(4/3)$-corrected KPZ    \\
        \hline
  $\beta_z$   &   $2.77\pm0.01$    &  $\frac{1}{6} \left(13+\sqrt{13}\right)=2.76759\ldots$  &  $ \frac{1}{6} \left(13+\sqrt{13}\right)=2.76759\ldots$  \\
  $\beta_y$   &    $1.90\pm0.01$   &   $\frac{1}{6} \left(7+\sqrt{13}\right)=1.76759\ldots$   & 
  $1+\frac{\sqrt{22}}{2 \left(\sqrt{13}-1\right)}=1.90008\ldots$ \\
  $\beta_x$   &     $1.19\pm0.01$  &  $1$   & $1+\frac{\sqrt{6}}{6 \left(\sqrt{13}-1\right)}=1.15668\ldots$   \\
  $\beta_w$   &    $ 1.99\pm0.01$  &  $1+\frac{\sqrt{6}}{\sqrt{13}-1}=1.94010\ldots$  &             
  $1+\frac{2 \sqrt{15}}{3(\sqrt{13}-1)}=1.99096\ldots$   \\
  $\beta_v$   &     $2.38 \pm0.03^{(*)}$  &  $1+\frac{2 \sqrt{3}}{\sqrt{13}-1}=2.32951\ldots$   &  
  $1+\frac{2 \sqrt{33}}{3(\sqrt{13}-1)}=2.46983\ldots$ \\
  $\beta_u$   &     $1.32\pm0.02$   &  $\frac{\sqrt{3}+\sqrt{6}-1}{\sqrt{13}-1}=1.22106\ldots$   &  
  $ \frac{2 \sqrt{15}+\sqrt{33}-3}{3 \left(\sqrt{13}-1\right)}=  1.34207\ldots$      \\
  
       \hline
    \end{tabular}
    \caption{\small Comparison of the numerical estimates for the various configuration exponents and their
    values predicted by the naive and by the $(4/3)$-corrected KPZ formulas. 
    \hspace{\textwidth}
    $\null^{(*)}$As explained in the text and in Figure~\ref{fig:Betacheck},
    a more reliable estimate is $\beta_v=2.42\pm 0.06$.}
    \label{table:exponents}
\end{table}

\begin{figure}[h]
  \centering
  \fig{.65}{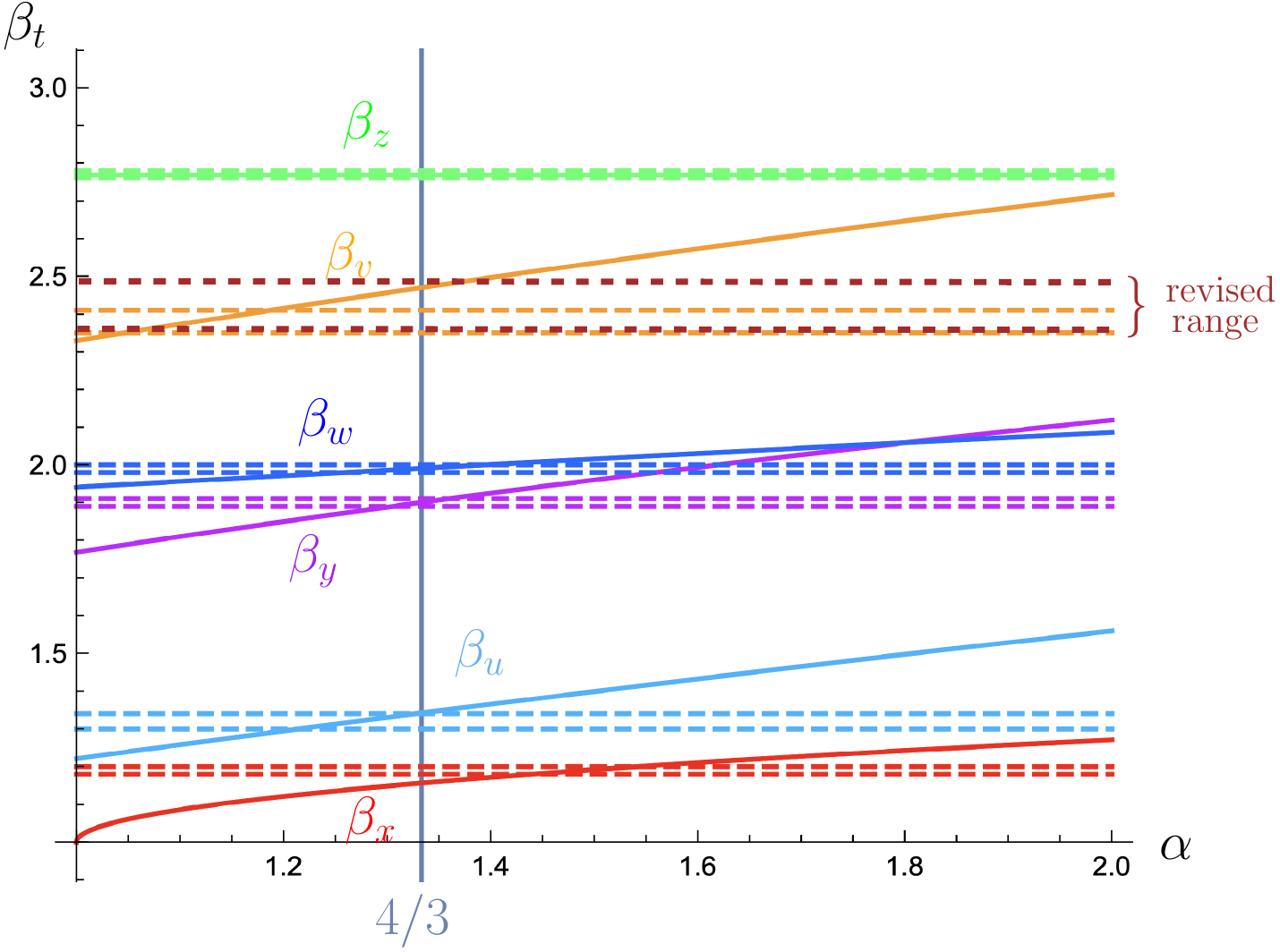}
  \caption{\small Comparison between the numerical estimates (dashed horizontal lines) of the various exponents and their ``$\alpha$-corrected'' KPZ 
  prediction (continuous lines) for $\alpha$ between $1$ and $2$. For $\beta_v$, we indicated in brown the revised extended range found
  in Figure~\ref{fig:Betacheck}. A reasonable matching is obtained for $\alpha \sim  4/3$.}
  \label{fig:alphavar}
\end{figure}

\bigskip
Let us now come to the prediction for the exponents $\beta_y,\beta_x,\ldots\beta_u$ themselves. 
As displayed in Table~\ref{table:exponents}, we find without any doubt that the ``naive'' prediction \eqref{eq:Delta0h}-\eqref{eq:KPZpred}
above \emph{does not match} with our numerical results. However, a reasonable agreement may again be recovered if,
as done in Section~\ref{sec:none} for $n=1$, we perform a modification of $\Delta_{\M}$ into
\begin{equation}
\Delta_{\M}:=\Delta(h^{(\alpha)}_{\M},-1)=\frac{\sqrt{1+12 h^{(\alpha)}_{\M}}-1}{\sqrt{13}-1}\ ,
\label{eq:Delta0halpha}
\end{equation}
with $h^{(\alpha)}_{\M}$ defined as in \eqref{eq:halphadef}, now for $n=0$, that is
\begin{equation}
h^{(\alpha)}_{\M}=\frac{\alpha}{24}\phi_1^2 +\frac{1}{8}\left(1-\delta_{\phi_2,0}\right)(\phi_2^2-1)\qquad
 \hbox{for}\quad \M=\phi_1 \A+\phi_2\btwo \ 
\label{eq:hM0alpha}
\end{equation}
and for a suitable choice of $\alpha$. 
Figure~\ref{fig:alphavar} displays the value of the ``$\alpha$-corrected'' KPZ prediction for
the various exponents for a varying value of $\alpha$ between $1$ and $2$. We see that, while the value $\alpha=1$ (naive KPZ prediction)
is clearly ruled out, a reasonable agreement may be obtained if we take $\alpha \sim  4/3$. The ``$(4/3)$-corrected'' KPZ 
predictions are listed in Table~\ref{table:exponents} for a direct comparison with numerics.

\section{Discussion}
\label{sec:discussion}
We have seen at the end of Section~\ref{sec:none} that for the FPL(1) model, the presence of a normalization factor $\alpha=9/8$ proved necessary in the Coulomb gas formula  \eqref{eq:halphadef} for $h^{(\alpha)}_{\M}$, in order to recover from the (inverse) KPZ formula the quantum gravity exponents $\Delta_{\ell}$ of \cite{K00} given
by  \eqref{eq:hlneufhuit}. Similarly, the numerical study of Sections~\ref{sec:numerics} and \ref{sec:comparison} showed that various numerical critical exponents associated with FPL(0) on random bicubic maps could be consistently obtained by KPZ via the introduction of a similar factor $\alpha\sim 4/3$ in \eqref{eq:halphadef}. We shall here try to give a possible meaning
to these observed values. 

\begin{figure}[h]
  \centering
  \fig{.9}{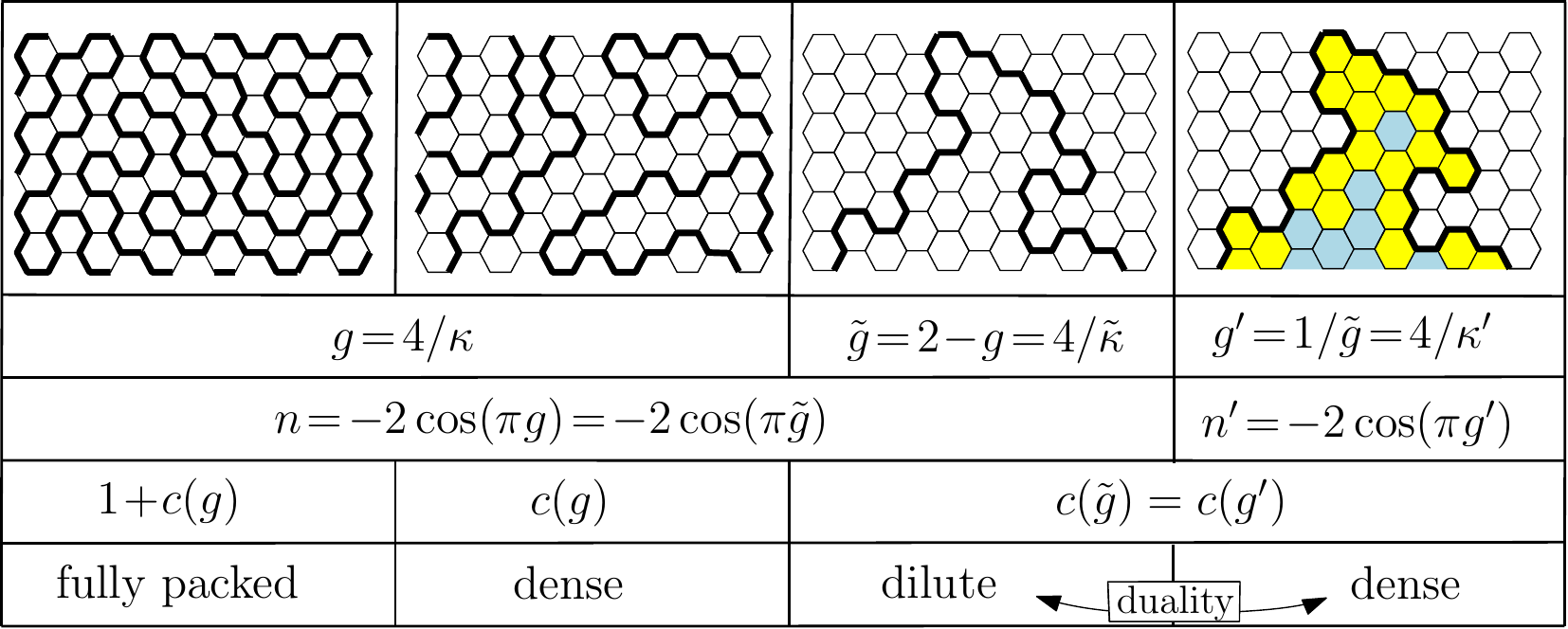}
  \caption{\small Illustration of the different loop models on the honeycomb lattice and their connections. We have $\frac{1}{2}\leq g\leq 1$,
  $1\leq \tilde{g}\leq \frac{3}{2}$ and $\frac{2}{3}\leq g'\leq 1$. For $n=0$, we have $g=\frac{1}{2}$ ($\kappa=8$), $\tilde{g}=\frac{3}{2}$ ($\tilde{\kappa}=\frac{8}{3}$)), 
  $g'=\frac{2}{3}$ ($\kappa'=6$), hence $n'=1$. For $n=1$, we have $g=\frac{2}{3}$ ($\kappa=6$), $\tilde{g}=\frac{4}{3}$ ($\tilde{\kappa}=3$)
  $g'=\frac{3}{4}$ ($\kappa'=16/3$), hence $n'=\sqrt{2}$. Top line from left to right, $(n=0, n'=1)$ case:  Hamiltonian walk, dense walk, SAW, SAW as the external perimeter of a percolation cluster.}
  \label{fig:diversg}
\end{figure}

We may rewrite 
\eqref{eq:halphadef} as 
\begin{equation}
\label{eq:halphadefbis}
h^{(\alpha)}_{\M}:=\frac{g'}{12}\phi_1^2 +\frac{g}{4}\left(1-\delta_{\phi_2,0}\right)\left(\phi_2^2-\left(1-g^{-1}\right)^2\right)\qquad
 \hbox{for}\quad \M=\phi_1 \A+\phi_2\btwo\ , 
\end{equation}
by distinguishing two coupling constants, 
\begin{equation}\label{eq:gg'}
g':=\alpha g, \quad g=1-e_0=\frac{1}{\pi} \arccos \left(-\frac{n}{2}\right)\ .
\end{equation}
This corresponds to decoupling the scales of the two fields $\psi_1$ and $\psi_2$ in the Coulomb gas action \eqref{eq:action}, and replacing there the Gaussian term by 
$ \frac{1}{3} g' (\nabla \psi_1)^2 + g (\nabla \psi_2)^2$. 
It is noteworthy that the \emph{Ansatz},
\begin{equation}\label{eq:Ansatz}
\alpha=\frac{1}{1-e_0^2}\ ,
\end{equation}
reproduces $\alpha=9/8$ for $n=1$ and $\alpha=4/3$ for $n=0$. In turn, it yields the coupling constant of the $\psi_1$ Gaussian free field,
\begin{equation}\label{eq:g'} 
g':=\frac{1}{1+e_0}\ .
\end{equation}
To the constrained $\mathrm{FPL}(n)$ model on the honeycomb lattice corresponds an unconstrained $\hbox{O}(n)$ model \cite{BN94,BSY94,KdGN96}, whose (stable) \emph{dense} critical phase has  the same CG coupling constant $ g=1-e_0$, with 
$1/2 \leq g \leq 1$ for $n=-2\cos (\pi g) \in [0,2]$, and a central charge 
\begin{equation}
c(g):=1-6\frac{(1-g)^2}{g} \ ,
\label{eq:cofg}
\end{equation} 
such that $ -2 \leq c(g) \leq 1$, see Figure~\ref{fig:diversg}.  
The associated (unstable) \emph{dilute} critical phase of the same $\hbox{O}(n)$ model has coupling constant $\tilde g:=1+e_0=2-g$, with $1\leq \tilde g \leq 3/2$  \cite{zbMATH03959008}, such that 
$n=-2\cos (\pi \tilde g)$, but with a different central charge, $c(\tilde g)=1-6(1-\tilde g)^2/\tilde g$, such that  $0 \leq c(\tilde g) \leq 1$. 

The coupling contant $g'$ of Equation~\eqref{eq:g'} then appears to be the \emph{dual} value of $\tilde g$, 
\begin{equation}\label{eq:duality}
g'=\frac{1}{\tilde g}=\frac{1}{2-g}\ .
\end{equation}
Because of its range, $2/3 \leq g'\leq  1$, this  CG coupling constant $g'$ corresponds to the \emph{dense} phase of \emph{another} $\hbox{O}(n')$ model, such that $n'=-2\cos (\pi g')$, but with the \emph{same} central charge as that of the dilute critical $\hbox{O}(n)$ model since, as easily checked, we have $c(g')=c(\tilde g)$. 
The geometrical interpretation of this  \emph{duality} is as follows \cite{Duplantier00,Duplantier03,MR2112128}. 
In the scaling limit, the loops of the dense $\hbox{O}(n')$ model are \emph{non-simple} random paths of Hausdorff dimension $D'=1+(2g')^{-1}$ \cite{zbMATH03959008}; their \emph{external perimeters} are \emph{simple} critical lines of the dilute critical $\hbox{O}(n)$ model, of Hausdorff dimension $\tilde D=1+(2\tilde g)^{-1}$. These Hausdorff dimensions thus satisfy the (super-)universal duality relation $(D'-1)(\tilde D-1)=1/4$ \cite{Duplantier00}, which can be directly obtained in the case of critical percolation \cite{PhysRevLett.83.1359}. 

This Coulomb gas $g \leftrightarrow 1/g$  ``electro-magnetic'' duality directly leads  to the $\kappa \leftrightarrow 16/\kappa$ duality of Schramm-Loewner evolution SLE$_\kappa$ \cite{Duplantier00,Duplantier03,MR2112128}.  The CG coupling constant spans the range $g\in [1/2,3/2]$ for the (dense and dilute) critical $\hbox{O}(n)$ model (for $n\geq 0$), and the SLE parameter the range $\kappa=4/g \in [8/3,8]$ for its scaling limit, the conformal loop ensemble $\mathrm{CLE}_\kappa$. When $\kappa \in (4,8]$, $\mathrm{SLE}_{\kappa}$ paths 
are non-simple \cite{MR2153402},  and their outer 
boundaries have been proven to be dual simple SLE$_{16/\kappa}$ paths, with $16/\kappa\in [2,4)$ \cite{MR2439609,dub_dual}.

\bigskip
To the fully-packed $\mathrm{FPL}(n\!=\!1)$ model considered in Section~\ref{sec:none} above corresponds an unconstrained dense $\hbox{O}(n\!=\!1)$ model with CG coupling constant $g=2/3$ and central charge $c_{\mathrm{dense}}(n\!=\!1)=0$, describing in particular critical percolation. The associated dilute phase has $\tilde g=4/3$, which is associated with the critical Ising model of central charge $c=1/2$. By duality, we find a second dense $\hbox{O}(n'\!=\!\sqrt{2})$ model with coupling constant $g'=3/4$, which also describes the Fortuin-Kasteleyn (FK) clusters of the critical $Q=2$ Potts model. In terms of $\mathrm{SLE}_\kappa$, critical percolation corresponds to  
$\kappa=4/g=6$ \cite{SMIRNOV2001239}, the critical Ising model to  $\tilde \kappa := 4/\tilde g=3$ \cite{zbMATH05808591}, while $Q=2$ FK clusters generate dual $\mathrm{SLE}_{16/3}$ random paths.

For the $\mathrm{FPL}(n\!=\!0)$ model describing Hamiltonian cycles and paths as studied in Sections~\ref{sec:numerics} and \ref{sec:comparison}, the corresponding unconstrained dense $\hbox{O}(n\!=\!0)$ model is that of dense polymers with $g=1/2$, $c_{\mathrm{dense}}(n\!=\!0)=-2$; the dilute phase is then naturally that of critical self-avoiding walks (SAW), i.e., dilute polymers with $\tilde g=3/2$ and $c=0$. By duality, the associated final dense $\hbox{O}(n'\!=\!1)$ model is that of percolation with $g'=2/3$ and $c=0$. In terms of $\mathrm{SLE}_\kappa$, one here goes from $\kappa=8$ to $\tilde \kappa= 8/3$ to $\kappa'= 6$, and it may be worth noting that in addition to critical percolation clusters \cite{PhysRevLett.83.1359,1999PhRvL..82.3940D}, 
planar Brownian loops also have the scaling limit of self-avoiding loops as external frontiers \cite{1999PhRvL..82..880D,zbMATH01690755,MR2112128bis}. 

To conclude this discussion, while formulas \eqref{eq:halphadefbis} \eqref{eq:gg'} and the Ansatz \eqref{eq:Ansatz} \eqref{eq:g'} \eqref{eq:duality} seem appealing, and lead to connections between various critical statistical models that show up when using KPZ relations  between fully-packed models on the honeycomb lattice and on random bicubic maps, it remains difficult at this stage to theoretically explain such apparent ``transmutations'' between models. Finally, let us remark that the observation that the usual KPZ relation \eqref{eq:Deltahc} or \eqref{eq:KPZdirect} fails for a family of exponents of the fully-packed models may be linked to a \emph{lack of independence} between some (constrained) configurations of the space-filling random paths and the random Liouville measure. Statistical independence is indeed crucial to the proof of that relation \cite{springerlink:10.1007/s00222-010-0308-1bis,2009arXiv0901.0277D}. The apparent enhancement of the effective coupling constant from $g$ to $g'=1/(2-g) \geq g$, for the extra Gaussian free field brought in by the full-packing condition, may reflect such a lack of independence. 

\bigskip 

\section*{Acknowledgements} We thank Ivan Kostov for interesting discussions. PDF is partially supported by the Morris and Gertrude Fine endowment and the NSF grant
DMS18-02044.
\appendix
\section{Hamiltonian paths on random cubic maps}
\label{appendix:cubic}
\def\cubic{\circ}
\def\ct{t^{\cubic}}
\def\cz{z^{\cubic}}
\def\cy{y^{\cubic}}
\def\cx{x^{\cubic}}
\def\cw{w^{\cubic}}
\def\cv{v^{\cubic}}
\def\cu{u^{\cubic}}
\def\ch{h^{\cubic}}
\def\cbeta{\beta^{\cubic}}
\def\cDelta{\Delta^{\cubic}}
\def\cmu{\mu^{\cubic}}
We discuss here briefly the problem of Hamiltonian path configurations defined on \emph{cubic} planar maps.
In practice, all the definitions given in Section~\ref{sec:enumerate} remain unchanged \emph{except that
we suppress the black/white colors of the vertices} and consequently forget about any bicoloration constraint whatsoever.
In this way, we define new numbers $\cz_N,\cy_N,\ldots,\cu_N$ which are the analogs for cubic maps of the numbers $z_N,y_N,\ldots,u_N$.
 Note that all the maps which we consider have by construction an even number of trivalent vertices\footnote{The number
 $V_3$ of trivalent vertices is even for all cubic maps with, possibly, an arbitrary number $V_2$ of bivalent defects and an \emph{even} number 
 $V_1$ of univalent ones since $3V_3=2E-2V_2-V_1$ where $E$ is the number of edges.} so that the meaning of $N$ is unchanged (e.g., there are
$2N$ trivalent vertices in configurations enumerated by $\cz_N$). We again define the exponential growth rate $\cmu$ and the configuration
exponents $\cbeta_t$ via the large $N$ behaviors
 \begin{equation}
 \ct_N\sim \hbox{const.}\ \frac{(\cmu)^{2N}}{N^{\cbeta_t}}
 \label{eq:tN0}
 \end{equation}
for the various quantities at hand. It is a simple exercise to obtain the exact expression
\begin{equation}
\cz_N=\sum_{k=0}^{N}\binom{2N}{2k} \hbox{Cat}_k\hbox{Cat}_{N-k}=\hbox{Cat}_N\hbox{Cat}_{N+1}\ ,
\end{equation}
where $\hbox{Cat}_N=\binom{2N}{N}/(N+1)$ is the $N^{\hbox{th}}$ Catalan number.
We similarly get
\begin{equation}
\begin{split} 
&\cy_N = 2^{2N}\hbox{Cat}_{N+2}\ ,\qquad \cx_N =2^{2N}\hbox{Cat}_{N}\ , \qquad 
\cw_N =(2N-1) \hbox{Cat}_{N-1}\hbox{Cat}_{N}\ ,\\ 
&\cv_N = \frac{1}{2}(N-1)\hbox{Cat}_N\hbox{Cat}_{N+1}\ ,\qquad
\cu_N =\frac{1}{4}(2N-1)(2N-2)\hbox{Cat}_{N-1}\hbox{Cat}_{N}\ . \\ 
\end{split}
\end{equation}
From these exact expressions, we immediately obtain that all these sequences have the same exponential growth rate
$(\cmu)^2=16$, while the configuration exponents read:
\begin{equation}
\cbeta_z=3\ , \qquad
\cbeta_y=\cbeta_x=\frac{3}{2}\ , \qquad
\cbeta_w=\cbeta_v=2\, \qquad
\cbeta_u=1\ .
\end{equation} 
These values match with the predictions
\begin{equation}
\begin{matrix}
\cbeta_z=2-\gamma^{\cubic}\ , \quad \hfill &\cbeta_y=1+2\cDelta_{\frac{3}{2}\A+\frac{1}{2}\btwo}-\gamma^{\cubic}\ , \quad \hfill&
\cbeta_x= 1+2\cDelta_{-\frac{1}{2}\A+\frac{1}{2}\btwo}-\gamma^{\cubic}\ , \hfill  \\ \\ \cbeta_w=1+2\cDelta_{\A}-\gamma^{\cubic} 
\ , \quad \hfill &\cbeta_v= 1+2\cDelta_{2\A}-\gamma^{\cubic}\ , \quad \hfill  &\cbeta_u= \cDelta_{2\A}+2\cDelta_{\A}-\gamma^{\cubic}\ , \hfill 
\end{matrix}
\end{equation}
upon taking $\cDelta_{\frac{3}{2}\A+\frac{1}{2}\btwo}=\cDelta_{-\frac{1}{2}\A+\frac{1}{2}\btwo}=-1/4$, $\cDelta_{\A}=\cDelta_{2\A}=0$
and $\gamma^{\cubic}=-1$.  The latter value is that predicted by the KPZ formula \eqref{eq:gammac} since, as discussed in Section~\ref{sec:bicubic}
$c=-2$ is the expected central charge when the FPL$(0)$ model is defined on cubic planar maps, and $\gamma(-2)=-1$.
As for the dressed dimensions $\cDelta$, their values do not depend on the component along the $\A$ direction, which is compatible with the fact that
we must set $\A=\boldsymbol{0}$: we are therefore left in practice with the two independent exponents $\cDelta_{\frac{1}{2}\btwo}=-1/4$ and
$\cDelta_{\boldsymbol{0}}=0$, corresponding respectively to the  1- and 2-leg watermelon exponents \cite{DK88}.  
From the KPZ formula  \eqref{eq:Deltahc} which, at $c=-2$, reduces to
\begin{equation}
\Delta(h,-2)=\frac{\sqrt{1+8\, h}-1}{2}
\end{equation}
we identify
\begin{equation}
 \cDelta_{\frac{1}{2}\btwo}=\Delta( \ch_{\frac{1}{2}\btwo},-2) \qquad \hbox{and}\qquad
\cDelta_{\boldsymbol{0}}=\Delta( \ch_{\boldsymbol{0}},-2)
\end{equation}
with the classical dimensions $\ch_{\frac{1}{2}\btwo}=-3/32$ and $\ch_{\boldsymbol{0}}=0$ \cite{DK88}. These values are two instances with $\phi_2=1/2$ and $0$ 
respectively of the general formula \eqref{eq:hM0}, which when $\A=\boldsymbol{0}$, translates into
\begin{equation}
\ch_{\phi_2\btwo}=\frac{1}{8}\left(1-\delta_{\phi_2,0}\right)(\phi_2^2-1)\ .
\end{equation}
To conclude, the obtained configuration exponents for the FPL$(0)$ model defined on cubic planar maps are exactly those 
predicted by the KPZ formulas. This confirms that the discrepancies with KPZ found in this paper for \emph{bicubic} maps are due to the existence
of the extra dimension (along $\A$) in the problem.   

\begin{table} 
\section{Enumeration results}
\label{appendix:results}
$\ $

    \centering
       \begin{tabular}{|rr|rr|}
        \hline
  $N$   &         $z_N$  &   $N$   &                       $z_N$   \\
        \hline
  1   &             2  &  15   &               1103650297320   \\
  2   &             8  &  16   &               9450760284100   \\
  3   &            40  &  17   &              81696139565864   \\
  4   &           228  &  18   &             712188311673280   \\
  5   &          1424  &  19   &            6255662512111248   \\
  6   &          9520  &  20   &           55324571848957688   \\
  7   &         67064  &  21   &          492328039660580784   \\
  8   &        492292  &  22   &         4406003100524940624   \\
  9   &       3735112  &  23   &        39635193868649858744   \\
 10   &      29114128  &  24   &       358245485706959890508   \\
 11   &     232077344  &  25   &      3252243000921333423544   \\
 12   &    1885195276  &  26   &     29644552626822516031040   \\
 13   &   15562235264  &  27   &    271230872346635464906816   \\
 14   &  130263211680  &  28   &   2490299924154166673782584   \\
        \hline
    \end{tabular}
    \vspace{-10pt}
    \caption{\small The number $z_N$ of Hamiltonian cycles on planar bicubic maps with $2N$ vertices, and a marked visited edge.}
    \label{table:base}
\bigskip
\centering
    \begin{tabular}{|rr|rr|}
        \hline
         $N$  &   $y_N$ & $N$ &$y_N$ \\
         \hline
         0  &        1  &   9  &         80576316  \\
         1  &        6  &  10  &        698497236  \\
         2  &       40  &  11  &       6125241762  \\
         3  &      286  &  12  &      54248935624  \\
         4  &     2152  &  13  &     484629868212  \\
         5  &    16830  &  14  &    4362375489180  \\
         6  &   135632  &  15  &   39532218657398  \\
         7  &  1119494  &  16  &  360393965832256  \\
         8  &  9421536  &      &                   \\
       \hline
    \end{tabular}
    \vspace{-10pt}
    \caption{\small The number $y_N$ of open Hamiltonian paths on planar bicubic with $2N+2$ vertices.}
    \label{table:semi-trivalent}
\bigskip
\centering
    \begin{tabular}{|rr|rr|}
        \hline
       $N$  &     $x_N$ & $N$  &            $x_N$  \\
         \hline
         0  &         1  &   9  &          56959872  \\
         1  &         4  &  10  &         512093760  \\
         2  &        24  &  11  &        4652471904  \\
         3  &       168  &  12  &       42641120752  \\
         4  &      1280  &  13  &      393739429376  \\
         5  &     10288  &  14  &     3659068137088  \\
         6  &     85776  &  15  &    34193890019424  \\
         7  &    734448  &  16  &   321103772899152  \\
         8  &   6416912  &  17  &  3028414925849920  \\
        \hline
    \end{tabular}
    \vspace{-10pt}
    \caption{\small The number $x_N$ of open Hamiltonian paths on planar bicolored maps with $2$ univalent vertices and $2N$ trivalent ones.}
    \label{table:semi}
\end{table}
\begin{table} 
    \centering
    \begin{tabular}{|rr|rr|}
        \hline
       $N$  &     $w_N$  & $N$  &             $w_N$  \\
        \hline
         1  &         1  &  10  &          29734848  \\
         2  &         4  &  11  &         251955792  \\
         3  &        22  &  12  &        2165922244  \\
         4  &       140  &  13  &       18848640980  \\
         5  &       972  &  14  &      165764482320  \\
         6  &      7160  &  15  &     1471222986648  \\
         7  &     55068  &  16  &    13162929589308  \\
         8  &    437692  &  17  &   118606870664836  \\
         9  &   3570100  &  18  &  1075505940036672  \\
    \hline
    \end{tabular}
    \vspace{-10pt}
       \caption{\small Table of $w_N$ where $2w_N$ is the number of cycles 
       visiting all the trivalent vertices on planar bicolored maps with $2$ (unvisited) univalent vertices and $2N$ trivalent ones.}
    \label{table:m+1m-1}
\bigskip
    \centering
        \begin{tabular}{|rr|rr|}
        \hline
       $ N$  &      $v_N$  &  $N$  &               $v_N$  \\
         \hline
         2  &          1  &  12  &          1996703248  \\
         3  &         10  &  13  &         17470889224  \\
         4  &         84  &  14  &        154096032108  \\
         5  &        682  &  15  &       1369014000682  \\
         6  &       5534  &  16  &      12242457072892  \\
         7  &      45330  &  17  &     110131946780584  \\
         8  &     375868  &  18  &     996123282195032  \\
         9  &    3155704  &  19  &    9054534704495656  \\
        10  &   26808852  &  20  &   82678808925578480  \\
        11  &  230230658  &  21  &  758122496862199740  \\
        \hline
    \end{tabular}
    \vspace{-10pt}
    \caption{\small The number $v_N$ of cycles 
       visiting all the trivalent vertices on planar bicolored maps with $2$ (unvisited) bivalent vertices and $2N$ trivalent ones.
       The bivalent vertices have necessarily different colors and the edges incident to the black one are distinguished.}
    \label{table:m+2m-2}
\bigskip
    \centering
      \begin{tabular}{|rr|rr|}
        \hline
        $N$  &    $u_N$  &   $N$  &            $u_N$  \\
         \hline
         2  &        1  &  10  &         47714564  \\
         3  &       10  &  11  &        439727448  \\
         4  &       90  &  12  &       4075738256  \\
         5  &      798  &  13  &      37971881232  \\
         6  &     7094  &  14  &     355404743524  \\
         7  &    63508  &  15  &    3340333168292  \\
         8  &   573056  &  16  &   31512818722844  \\
         9  &  5210640  &  17  &  298306803039300  \\
        \hline
    \end{tabular}
    \vspace{-10pt}
    \caption{\small Table of $u_N$ where $4u_N$ is the number of cycles 
       visiting all the trivalent vertices on planar bicolored maps with $1$ (unvisited) bivalent black vertex with distinguished incident edges, 
       $2$ (unvisited) univalent vertices and $2N$ trivalent ones.}
    \label{table:m+2m-1m-1}
\end{table}

\begin{table} 
\end{table} 
\newpage
\bibliographystyle{unsrt}
\bibliography{bicubicFPL}
\end{document}